\documentclass[aps,prl,twocolumn,superscriptaddress,10pt,floatfix]{revtex4-2}

\usepackage[utf8]{inputenc}
\usepackage[T1]{fontenc}
\usepackage[english]{babel}
\usepackage{amsmath}
\usepackage{amssymb}
\usepackage{graphicx}
\usepackage{xcolor}
\usepackage[colorlinks=true, linkcolor=blue, urlcolor=blue, citecolor=blue]{hyperref}

\begin{document}

\title{Hidden entropic order of emergent charges in a classical Amp\`ere spin liquid}

\author{H. Diego Rosales}
\affiliation{Instituto de F\'isica de L\'iquidos y Sistemas Biol\'ogicos, CONICET, Facultad de Ciencias Exactas, Universidad Nacional de La Plata, 1900 La Plata, Argentina}
\affiliation{Departamento de Ciencias B\'asicas, Facultad de Ingenier\'ia, UNLP, La Plata, Argentina}

\author{Emma M. Bizot}
\affiliation{Laboratoire de Physique Th\'eorique, Universit\'e de Toulouse, CNRS, France}
\author{Flavia A. G\'omez Albarrac\'in}
\affiliation{Instituto de F\'isica de L\'iquidos y Sistemas Biol\'ogicos, CONICET, Facultad de Ciencias Exactas, Universidad Nacional de La Plata, 1900 La Plata, Argentina}
\affiliation{Departamento de Ciencias B\'asicas, Facultad de Ingenier\'ia, UNLP, La Plata, Argentina}

\author{Pierre Pujol}
\affiliation{Laboratoire de Physique Th\'eorique, Universit\'e de Toulouse, CNRS, France}

\date{\today}

\begin{abstract}
Classical spin liquids are characterized by an extensive number of ground-state configurations. Coulomb spin liquids have, in this ground-state manifold, a divergence-free local constraint, which gives dipolar correlations and pinch points in the structure factor. Recently, models with a complementary ground-state constraint, defined by a zero-circulation constraint (Amp\`ere's law), have been proposed. While classical Coulomb spin liquids forbid emergent charges, their Amp\`ere counterparts leave the charges unconstrained. Here we study an Ising Amp\`ere model on the octochlore lattice. The ground states obey a Gauss law for an antisymmetric rank-two field, i.e., three coupled two-dimensional Gauss laws, which are expected to produce pinch points in the structure factor. Nevertheless, maximal charges, vertices with all six arrows in or out, spontaneously accumulate on one of the two sublattices of the cubic lattice, with disordered signs. This order has a purely entropic origin; it washes out the pinch-point singularities at large scales, produces no magnetic Bragg peaks, and coexists with an extensive entropy. The system enters into this sublattice long-range ordered state through a continuous transition in what seems to be the three-dimensional Ising universality class. Moreover, the state has no 
magnetic order; it is even under time reversal: it is a hidden, entropic order within the Amp\`ere manifold. Thus, a system whose Gauss law would predict an algebraic spin liquid becomes, through a non-magnetic ordering transition, a partially ordered short-range spin liquid \footnote{Although a lattice symmetry is broken, as this is a classical system, and in contrast to quantum systems, we keep the terminology of spin liquid because it is a state with no magnetic order and extensive entropy.}. Our study is based on a combination of the constraint-vector theory, exact bounds and enumeration, and Monte Carlo, Wang-Landau, and parallel-tempering simulations.
\end{abstract}

\maketitle


\emph{Introduction.}---Classical spin liquids are frustrated magnets whose ground states form an extensive manifold that the system keeps exploring as $T\to0$~\cite{OBD,Wannier,Anderson_Ice,Youngblood_Ice,Garanin_Ice}. A systematic way to build them is to write the Hamiltonian as a sum of squares of local ``plaquette'' spins, $\mathcal{H}=\frac J2\sum_p\boldsymbol{\mathcal S}_p^2$, so that the ground states satisfy $\boldsymbol{\mathcal S}_p=0$ everywhere~\cite{benton2021topological,davier2023combined,Davier_Pinch_Lines}. When this constraint is a zero divergence, the system is a Coulomb spin liquid (CSL): a lattice Gauss law gives dipolar correlations and pinch points in the structure factor~\cite{henley2005power,Isakov_Dipolar,davier2023combined,Yan1,Yan2}, and the violations of the constraint are emergent charges with Coulomb interactions.

Recently, Chung and Gingras~\cite{chung20252} and Rougemaille and co-workers~\cite{rougemaille2025ampere} introduced the complementary class of \emph{Amp\`ere} spin liquids (ASL). Ising spins on the links of a lattice are read as arrows, and the ground-state constraint is a vanishing \emph{circulation} of the arrow field: the field is curl-free rather than divergence-free. These models also display pinch points, which on the pyrochlore lattice coexist with sharp, Bragg-like peaks~\cite{chung20252}. A natural question is then what role the charges play in an ASL. Since the Hamiltonian constrains the curl and not the divergence of the arrow field, the charges $Q=\nabla\cdot\mathbf m$ are free in the ground states.

\begin{figure}[t]
	\centering
	\includegraphics[width=0.95\linewidth]{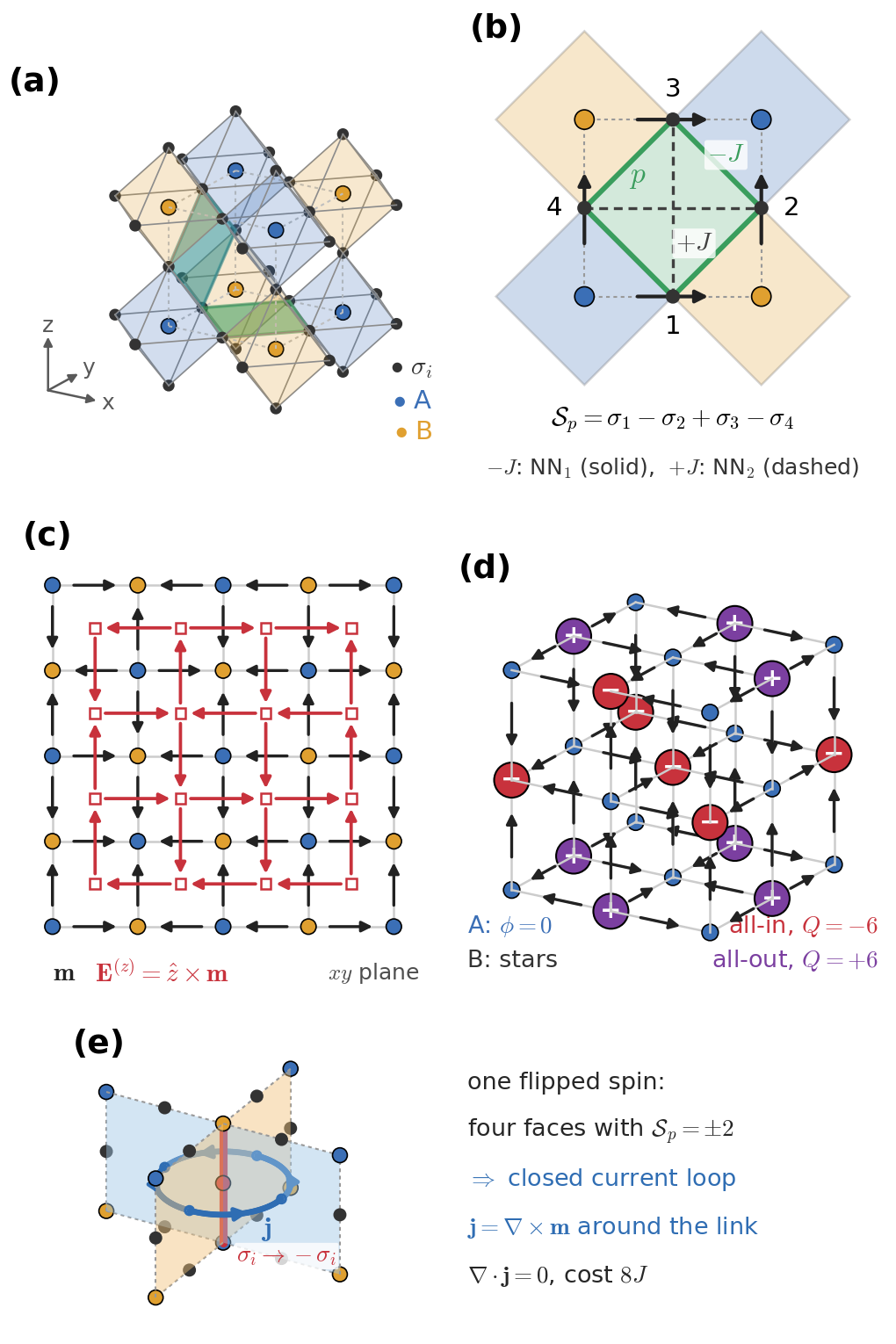}
	\caption{(a) Octochlore lattice: Ising spins $\sigma_i$ (black) at the vertices of corner-sharing octahedra centred on the vertices of a cubic lattice (sublattices $A$, blue, and $B$, orange); two plaquettes are shown. (b) An $xy$ plaquette from above, with a configuration with $\mathcal S_p=0$ (arrows: $\mathbf m$). (c) A ground state in an $xy$ plane: arrows $\mathbf m$ (black) and $\mathbf E^{(z)}=\hat z\times\mathbf m$ (red) on the dual links, with zero flux at every face center. (d) Star order: $\phi=0$ on $A$ and $\phi=\pm1$ on $B$ make every $B$ vertex a star, all-in (red) or all-out (purple). (e) Smallest defect: one flipped spin (red) and its current loop $\mathbf j=\nabla\times\mathbf m$ (blue) through four faces.}
	\label{fig:lattice}
\end{figure}

In this Letter, we study the analogous model on the octochlore lattice, where Ising spins sit on the links of a cubic lattice. We first make explicit the Gauss law hidden behind the Amp\`ere constraint: in two dimensions the ASL is exactly equivalent to a CSL, and in three dimensions, on the octochlore lattice, we show that it becomes a Gauss law for an antisymmetric rank-two tensor, i.e., three inequivalent coupled two-dimensional Gauss laws, which account for the pinch points. We then show that the Ising ASL on this lattice is unstable to a \emph{hidden} order: below $T_c=1.6309(15)J$ the maximal charges, ``stars'' with all six arrows all-in or all-out (AIAO), accumulate on one of the two sublattices of the cubic lattice with random signs. The order is purely entropic and gives no magnetic Bragg peak; its signature in the magnetic structure factor is a set of strong diffuse maxima. The pinch-point pattern, in turn, is shaped by two mechanisms: above $T_c$ it is broadened by plaquette defects, and below $T_c$ the star order suppresses the long-wavelength longitudinal fluctuations, washing out the pinch-point singularities. The low-temperature phase has thus short-range magnetic correlations (this scenario is reminiscent of the ``order with disorder'' of Azaria et al. in which one sublattice orders while the other remains disordered down to $T=0$~\cite{Azaria1987}; here the ordered quantity is the magnitude of the charges rather than the spins). The transition is continuous and consistent with the three-dimensional (3D) Ising class. These results are obtained with the constraint-vector (projector) formalism, an exact bound on the residual entropy and the exact enumeration of a small system, and three independent simulation methods: annealing Monte Carlo (MC) with a zero-energy cluster move, Wang-Landau (WL) sampling of the density of states and of the joint distribution of energy and order parameter, and parallel tempering (PT) up to $N\simeq10^5$ spins, analyzed by multiple-histogram reweighting and finite-size scaling (FSS). Details and complementary results are given in the Supplemental Material (SM).

\emph{Two dimensions: Amp\`ere = Coulomb.}---Consider arrows on the links of the square lattice, with the vertices split into sublattices $A$ and $B$ and the convention $\sigma=+1$ ($-1$) for an arrow from $A$ to $B$ (from $B$ to $A$)~\cite{rougemaille2025ampere}. With $\mathbf m$ the arrow field, the Hamiltonian $\mathcal{H}=\frac J2\sum_p\Gamma_p^2$ penalizes the circulation $\Gamma_p=\oint_p\mathbf m\cdot d\mathbf l$ around each plaquette, so the ground states are curl-free, $\partial_xm_y-\partial_ym_x=0$. Rotating the arrows by $90^\circ$,
\begin{equation}
	E_\mu=\epsilon_{\mu\nu}m_\nu ,
	\label{eq:E2D}
\end{equation}
turns the Amp\`ere law into a Gauss law, $\nabla\cdot\mathbf E=0$, for arrows living on the dual lattice. In terms of Ising variables this is the redefinition $\tilde\sigma=\epsilon\,\sigma$ ($\epsilon=\pm1$ on vertical/horizontal links), which maps the model exactly onto the antiferromagnetic Ising model on the checkerboard lattice, a well-known CSL~\cite{davier2023combined} (SM, Sec.~\ref{sm:2D}). In two dimensions, therefore, Amp\`ere and Coulomb spin liquids are the same object, seen from the lattice or from its dual.

\emph{Octochlore lattice and rank-two Gauss law.}---In three dimensions we place Ising spins on the $N=3L^3$ links of a periodic $L^3$ cubic lattice [the octochlore lattice of corner-sharing octahedra, Fig.~\ref{fig:lattice}(a)], with $m_i=\eta_i\sigma_i$ along the link and $\eta_i=\pm1$. The plaquettes are the faces of the cubes, in their three orientations $xy$, $xz$ and $yz$ [Fig.~\ref{fig:lattice}(b)], and
\begin{equation}
	\mathcal{H}=\frac J2\sum_p\mathcal S_p^2,\qquad \mathcal S_p=\sigma_1-\sigma_2+\sigma_3-\sigma_4 ,
	\label{eq:H}
\end{equation}
with the spins numbered cyclically around each face. Expanding $\mathcal S_p^2$, $\mathcal{H}$ is an Ising model with a ferromagnetic coupling $-J$ between nearest neighbors (NN$_1$, the sides of each plaquette, i.e., the edges of the octahedra) and an antiferromagnetic coupling $+J$ between the parallel spins across each face (NN$_2$, its diagonals) [Fig.~\ref{fig:lattice}(b)]. In the notation of Ref.~\cite{stern2026fracton}, this corresponds to $J_1=-J$, $J_{2b}=+J$ and $J_{2a}=0$; these couplings arise from an anisotropic exchange between the moment components transverse to each bond (SM, Sec.~\ref{sm:exchange}). Up to a sign, $\mathcal S_p$ is the circulation of $\mathbf m$ around $p$, and the ground states have $\nabla\times\mathbf m=0$ on every face. Excitations are faces with $\mathcal S_p\neq0$. Their circulations define a current $\mathbf j=\nabla\times\mathbf m$ on the links of the dual lattice, and since the circulations around each cube add up to zero (see below), $\nabla\cdot\mathbf j=0$: the defects are closed current loops rather than point charges~\cite{rougemaille2025ampere}. The smallest one is a single flipped spin, encircled by a loop through its four faces at a cost of $8J$ [Fig.~\ref{fig:lattice}(e)].
Note that, with periodic boundary conditions and at temperatures low enough that defects are absent, the states of the system split into topological sectors labeled by the circulation of $\mathbf m$ along a loop winding once around the system in the $x$, $y$ or $z$ direction; since $\nabla\times\mathbf m=0$, this circulation does not depend on the chosen loop.  On the cubic lattice, a curl-free field is a gradient, $m_{(\mathbf r,\mu)}=\phi(\mathbf r+\hat e_\mu)-\phi(\mathbf r)$, with $\phi\in\mathbb Z$ a height on the vertices and $|\Delta\phi|=1$ on every link. 

The two-dimensional construction can now be applied in each of the three families of planes. In the planes of normal $\hat e_\nu$ we rotate the arrows, $\mathbf E^{(\nu)}=\hat e_\nu\times\mathbf m$, and draw them on the dual links through the face centers [Fig.~\ref{fig:lattice}(c)]. Collecting the three rotated fields as the columns of a matrix gives the antisymmetric tensor associated to a plaquette:
\begin{equation}
	E_{\mu\nu}=\epsilon_{\mu\nu\rho}\,m_\rho ,\qquad \partial_\mu E_{\mu\nu}=0 .
	\label{eq:gauss}
\end{equation}
Each column $\nu$ of $E_{\mu\nu}$ is a vector lying in the planes perpendicular to $\hat e_\nu$ (its $\nu$ component vanishes), and $\partial_\mu E_{\mu\nu}=0$ is its two-dimensional Gauss law at the centers of the faces of normal $\hat e_\nu$. Equation~(\ref{eq:gauss}) is thus a compact notation for three two-dimensional Gauss laws, one per family of planes. They are not independent: each spin belongs to two families of planes, so each $m_\rho$ enters two columns (e.g.\ $E_{yz}=m_x$ and $E_{zy}=-m_x$), and the antisymmetry implies the identity $\partial_\nu\partial_\mu E_{\mu\nu}\equiv0$, i.e., only two of the three laws are independent. On the lattice this is the statement that the six circulations around a cube add up to zero.

The ASL is therefore described by a two-form (antisymmetric) $U(1)$ gauge field, as found for the pyrochlore ASL in Ref.~\cite{chung20252}, and not by a symmetric rank-two field such as those associated with the merging of pinch points in Ref.~\cite{davier2023combined}. 

\begin{figure}[t]
	\centering
	\includegraphics[width=\linewidth]{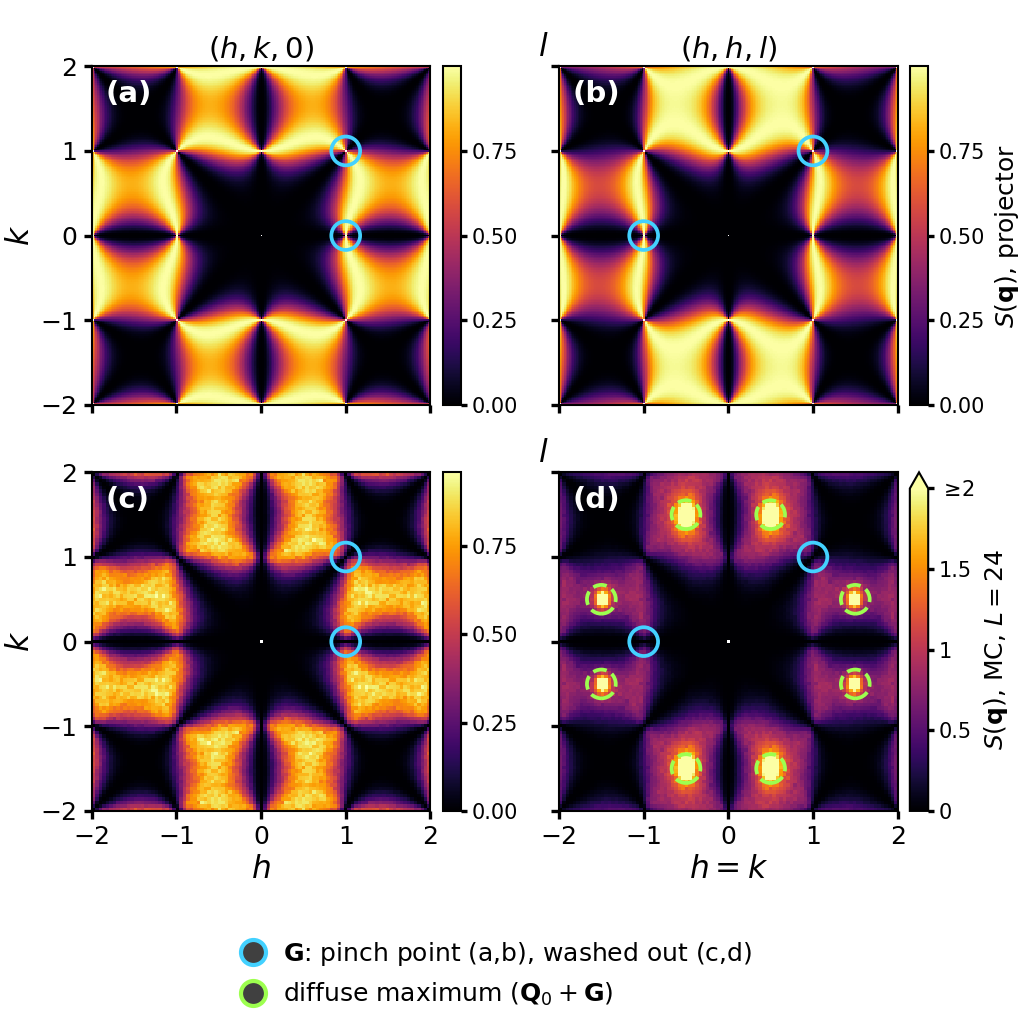}
	\caption{Neutron structure factor of the Ising Amp\`ere model on the octochlore lattice in the $(h,k,0)$ (left) and $(h,h,l)$ (right) planes, $\mathbf q=2\pi(h,k,l)$. (a,b) Projector, Eq.~(\ref{eq:projector}), with the polarization factor. (c,d) Annealing MC, $L=24$, $T=0.28J$, 500 configurations; in (d) the color scale is cut at $S=2$ to make the pinch-point pattern visible. Solid-line circles mark the angular pinch-point structure at some reciprocal-lattice vectors $\mathbf G$, present in both; dashed circles mark the additional diffuse maxima of MC at $\mathbf Q_0+\mathbf G$, $\mathbf Q_0=(\pi,\pi,\pi)$, e.g.\ $(h,l)=(\pm\frac12,\pm\frac32)$ and $(\pm\frac32,\pm\frac12)$, whose height ($S\simeq7$, beyond the bound $1$ of the projector) and width do not depend on $L$ (SM, Sec.~\ref{sm:Sq}). The pinch points are broadened by plaquette defects above $T_c$, and below it the star order (flat height field) washes out their singularity at large scales, while the angular structure remains (see main text).}
	\label{fig:Sq}
\end{figure}

In Fourier space Eq.~(\ref{eq:gauss}) reads $\mathbf q\times\mathbf m=0$: the arrow field is longitudinal, complementary to the transverse field of a CSL. The constraint-vector method~\cite{henley2005power,benton2021topological,davier2023combined} makes this quantitative on the lattice: labelling the three families of faces by their normal $\rho$, the constraint vectors are
\begin{equation}
	(\mathbf L_\rho)_\mu=2i\,\epsilon_{\rho\mu\nu}\,s_\nu ,
	\label{eq:L}
\end{equation}
with $s_\mu=\sin(q_\mu/2)$, e.g.\ $\mathbf L_z=2i(s_y,-s_x,0)$ for the $xy$ faces; $\mathbf L_\rho\!\cdot\!\mathbf m=2i(\mathbf m\times\mathbf s)_\rho$ is the lattice curl. They obey $\sum_\rho s_\rho\mathbf L_\rho=0$ (the Fourier form of the dependence just discussed), so that the matrix $\mathbf M=[\mathbf L_x,\mathbf L_y,\mathbf L_z]$, with the $\mathbf L_\rho$ as columns, has rank two. The correlations are given by the projector on the allowed space~\cite{henley2005power}, $\big\langle m_\mu(-\mathbf q)\,m_\nu(\mathbf q)\big\rangle\propto\Pi_{\mu\nu}(\mathbf q)$ with 
\begin{equation}
	\Pi=1-\mathbf M\big(\mathbf M^\dagger\mathbf M\big)^{+}\mathbf M^\dagger=\frac{\mathbf s\,\mathbf s^{T}}{|\mathbf s|^2},
	\label{eq:projector}
\end{equation}
which has pinch points at the reciprocal-lattice vectors $\mathbf G$ (SM, Sec.~\ref{sm:constraint}). As in spin ice~\cite{Isakov_Dipolar,fennell2009magnetic}, we compute the magnetic structure factor of the physical moments $\boldsymbol\mu_i=\sigma_i\eta_i\hat e_{\mu_i}$ (the arrows $\mathbf m_i$), with the polarization factor of unpolarized neutrons,
\begin{equation}
	S(\mathbf q)=\frac1N\sum_{i,j}\big\langle\boldsymbol\mu_i^{\perp}\cdot\boldsymbol\mu_j^{\perp}\big\rangle\,e^{i\mathbf q\cdot(\mathbf r_i-\mathbf r_j)},\qquad \boldsymbol\mu^\perp=\boldsymbol\mu-(\boldsymbol\mu\cdot\hat{\mathbf q})\hat{\mathbf q},
	\label{eq:Sq_def}
\end{equation}
with $\mathbf r_i$ the link midpoints; the projector prediction follows from Eq.~(\ref{eq:projector}) (SM, Sec.~\ref{sm:constraint}). Figure~\ref{fig:Sq} compares the two. The angular (bow-tie) pinch-point pattern is reproduced, but MC shows in addition sharp maxima at $\mathbf Q_0+\mathbf G$, $\mathbf Q_0=(\pi,\pi,\pi)$, with intensity $S\simeq7$, beyond the projector bound $S\le1$. Their height and width do not depend on $L$ ($L=8$--$24$), so they are not Bragg peaks (SM, Sec.~\ref{sm:Sq}), and they appear at $T_c$. They signal the transition discussed next. The constraint fixes the angular structure of $S(\mathbf q)$, while its intensity near $\mathbf G$ depends on the amplitude of the fluctuations, which is shaped by two mechanisms. Above $T_c$, plaquette defects (current loops, Fig.~\ref{fig:lattice}(e); a fraction $\sim0.1$ of the faces at $T_c$, SM) relax the curl-free constraint and broaden the pinch points, as monopoles do in spin ice~\cite{fennell2009magnetic}. Below $T_c$, the star order flattens the height field and suppresses the longitudinal fluctuations at large scales, leaving only short-range magnetic correlations: the angular structure survives, but the pinch-point singularity is washed out. The amplitudes of $S(\mathbf q)$ around $\mathbf G$ confirm both mechanisms and show that the bow tie keeps a finite width at the temperatures studied (SM, Fig.~\ref{fig:sm_pinch}). The bow-tie pattern is therefore most pronounced around $T_c$.

%

\emph{Free charges and residual entropy.}---At each vertex we define the charge $Q_{\mathbf r}=\nabla\cdot\mathbf m=\eta_{\mathbf r}\sum_{i\in\mathbf r}\sigma_i\in\{0,\pm2,\pm4,\pm6\}$, which the Amp\`ere Hamiltonian does not constrain, and call \emph{stars} the vertices with $|Q|=6$. A star can be reversed at no energy cost (in terms of the heights, $\phi\to\phi\pm2$ at that vertex); we use this ``star move'' in all our simulations to keep them ergodic inside the ground-state manifold, within a given winding sector.

\begin{figure}[t]
	\centering
	\includegraphics[width=0.75 \linewidth]{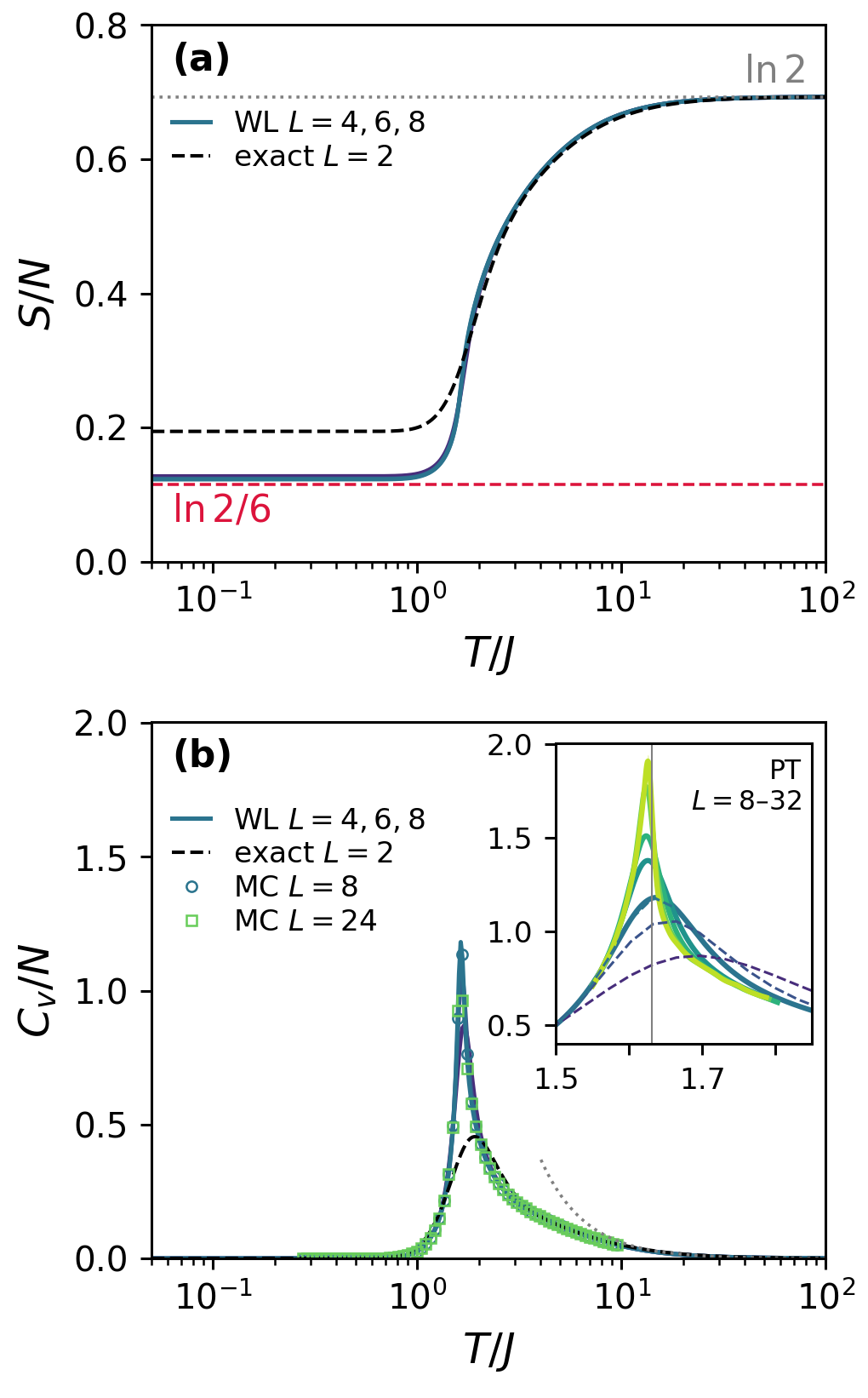}
	\caption{Thermodynamics of the Ising octochlore ASL from three methods. (a) Entropy per spin from WL ($L=4$--$8$, solid) and exact enumeration ($L=2$, dashed); dotted: the paramagnetic value $\ln2$; dashed red: the exact bound $\ln(2)/6$ from the star family of Fig.~\ref{fig:lattice}(d) (the improved bound of Eq.~(\ref{eq:bound}) is $0.1191$). (b) Specific heat per spin from WL (solid), exact $L=2$ (dashed) and annealing MC ($L=8$, $24$; symbols); dotted: the exact high-temperature limit $6J^2/T^2$. Inset: PT for $L=8$--$32$ (solid, dark to light) and WL (dashed) near $T_c$ (vertical line).}
	\label{fig:thermo}
\end{figure}

We can use the height representation as a tool to give an exact lower bound on the residual entropy $S_0$, i.e., on the number of ground states. This is done by choosing a globally flat configuration for the field by fixing $\phi=0$ on all vertices of $A$, and each vertex of $B$ can independently take $\phi=\pm1$ [Fig.~\ref{fig:lattice}(d)]: these $2^{L^3/2}$ ground states have \emph{all} $B$ vertices as stars, with random signs, and give $S_0/N\ge\ln(2)/6\simeq0.1155$. The bound can be further improved: an $A$ vertex whose six neighbours all have height $+1$ (respectively, $-1$ can also take $\phi=+2$ (respectively, $-2$, at no energy cost (a star move), so each sign configuration $s$ of $B$ yields $2^{n_A(s)}$ ground states, $n_A$ being its number of $A$ stars. Since $\langle n_A\rangle=L^3/64$ for random signs, convexity gives $\sum_s2^{n_A(s)}\ge2^{L^3/2}\,2^{L^3/64}$, i.e.,
\begin{equation}
S_0/N\ \ge\ \tfrac{33}{32}\,\tfrac{\ln2}{6}\simeq0.1191 .
\label{eq:bound}
\end{equation}
We computed the entropy from the WL density of states $g(E)$~\cite{WangLandau2001,BelardinelliPereyra2007}, validated against the exact enumeration of the $2^{24}$ states of $L=2$ [Fig.~\ref{fig:thermo}(a)]. $S(T)$ decreases from $\ln2$ to a plateau below $T\simeq J$, with $S_0/N=0.1273$, $0.1237$ and $0.1230$ for $L=4$, $6$, $8$, and $S_0/N=0.1223(1)$ extrapolated to $L\to\infty$ by fitting the entropy per spin, $s_0(L)\equiv S_0(L)/N=s_0^\infty+a/L^3$ (SM, Sec.~\ref{sm:thermo}). The residual entropy of the ASL is thus only $2.7\%$ above the bound~(\ref{eq:bound}), which suggests that typical ground states resemble the family in which all vertices of one sublattice are stars. Across the drop of $S(T)$ the specific heat has a single, sharp maximum [Fig.~\ref{fig:thermo}(b)] on which all methods agree: annealing MC, WL ($L\le8$) and PT ($L=8$--$32$), with $C_v^{\max}/N=1.18$, $1.38$, $1.51$, $1.77$ and $1.90$ for $L=8$, $12$, $16$, $24$, $32$ (SM, Sec.~\ref{sm:thermo}).
%

\begin{figure}[t]
	\centering
	\includegraphics[width=\columnwidth]{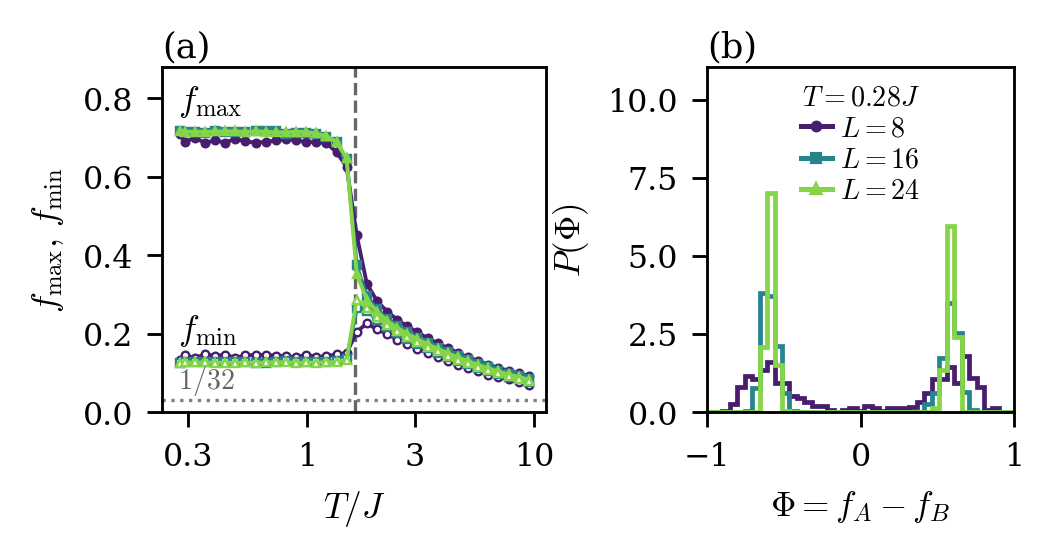}
	\caption{Star-charge order. (a) Fraction of stars ($|Q|=6$) on the sublattice with more ($f_{\max}$, upper branch) and fewer ($f_{\min}$, lower branch) stars, from annealing MC ($L=8$, $16$, $24$; 500 configurations per temperature; error bars: bootstrap over copies); dotted: $1/32$ (random arrows, $T\to\infty$); dashed: $T_c=1.631J$. Above $T_c$ the two sublattices are equivalent and the branches merge; below $T_c$ they saturate at $f_{\max}\simeq0.71$ and $f_{\min}\simeq0.13$, averages over the ground-state manifold. Inset: the same fractions from PT near $T_c$ ($L=8$--$32$, dark to light; solid $f_{\max}$, dashed $f_{\min}$). (b) Distribution of $\Phi=f_A-f_B$ at the lowest temperature: bimodal, each system choosing sublattice $A$ or $B$.}
	\label{fig:order}
\end{figure}

\emph{Hidden order of star charges.}---The proximity of $S_0$ to the bound suggests that, in typical ground states, the stars occupy one of the sublattices, and motivates the $\mathbb Z_2$ order parameter
\begin{equation}
	\Phi=f_A-f_B ,
	\label{eq:Phi}
\end{equation}
with $f_{A,B}$ the fractions of stars on the two sublattices. Figure~\ref{fig:order} shows that $\Phi$ orders. Below $T_c$ one sublattice carries $f\simeq0.71$ stars and the other $f\simeq0.13$ (to be compared with $1$ and $1/32$ in the saturated case): typical ground states are close to, but not in, that family of saturated states, and Monte Carlo sampling of a restricted family of ground states reproduces both values (SM, Sec.~\ref{sm:order}). The order parameter reaches the value $\langle|\Phi|\rangle\simeq0.58$ independent of $L$, the Binder cumulant reaches $2/3$, and the distribution of $\Phi$ is bimodal, each system choosing $A$ or $B$. Defects, faces with $\mathcal S_p\neq0$, form closed current loops costing at least $8J$; their density $\rho_{\rm def}$ is $\sim0.1$ at $T_c$, and they are negligible below $T\simeq0.45J$ (SM, Fig.~\ref{fig:sm_PT_obs}(c)). Below that temperature, where defects are absent, $\langle|\Phi|\rangle$ reaches a plaetau: the order takes place within the ground-state manifold itself and has a purely entropic origin. The selection itself involves no energy scale: typical states drawn uniformly from the ground-state manifold are already ordered. Temperature enters only through the defects, which take the system out of the ground-state manifold and restore the symmetry; this suggests that $T_c$ is controlled by their density ($\rho_{\rm def}\sim0.1$ at $T_c$).

The order is hidden in two respects. The fact that there is no magnetic order implies that the signed staggered charge $\sum_{\mathbf r}\eta_{\mathbf r}Q_{\mathbf r}=2\sum_i\sigma_i$ is not ordered: the signs of the stars have no long-range order, the order is in $|Q|$. The sublattice order is also reflected in correlations of the local pair observable $P({\mathbf r})=\frac{1}{15}\sum_{i<j\in {\mathbf r}}\sigma_i\sigma_j=(Q_{\mathbf r}^2-6)/30$, which equals unity for both all-in and all-out stars. Below $T_c$, its correlations approach distinct long-distance plateaus on the star-rich and star-poor sublattices, consistent with long-range sublattice order (SM, Fig.~\ref{fig:sm_pair_correlations}).

Since $\mathbf m$ remains a gradient, the angular structure of Eq.~(\ref{eq:gauss}) survives the ordering, while the star order washes out the pinch-point singularity at large scales (Fig.~\ref{fig:Sq}). The diffuse maxima at $\mathbf Q_0+\mathbf G$, which appear at $T_c$, are the fingerprint of the star order in the structure factor. Uncorrelated star signs on a fully occupied sublattice would give $S(\mathbf Q_p)=\frac23[3-(\hat{\mathbf q}\cdot\mathbf s)^2]\simeq1.9$ at $\mathbf Q_p=2\pi(\frac12,\frac12,\frac32)$. The counting of signed stars (see SM, Sec.~\ref{sm:order}), shows why the signs are correlated at short distance: sign configurations with locally aligned signs carry more ground states. Monte Carlo sampling of this restricted ground-state ensemble gives $S(\mathbf Q_p)\simeq7$, with no significant size dependence over the sizes studied, accounting for the maxima observed in the full MC simulations.

\begin{table}[b]
\caption{Star-order transition from PT ($L=8$--$32$; statistical errors), compared with the 3D Ising~\cite{Hasenbusch2010} class.}
\label{tab:exponents}
\begin{ruledtabular}
\begin{tabular}{lcc}
 & This work & 3D Ising\\
\hline
$T_c/J$ & $1.6309(15)$ & -- \\
$1/\nu$ & $1.559(13)$ & $1.587$ \\
$\gamma/\nu$ [$\chi'(T_c)$] & $1.986(9)$ & $1.964$ \\
$\gamma/\nu$ [$\chi_{\max}$] & $1.883(14)$ & $1.964$ \\
$\beta/\nu$ & $0.484(5)$ & $0.518$ \\
$U^*$ & $\simeq0.38$ & $0.4655$ \\
\end{tabular}
\end{ruledtabular}
\end{table}
\begin{figure*}[t]
	\centering
	\includegraphics[width=0.9 \textwidth]{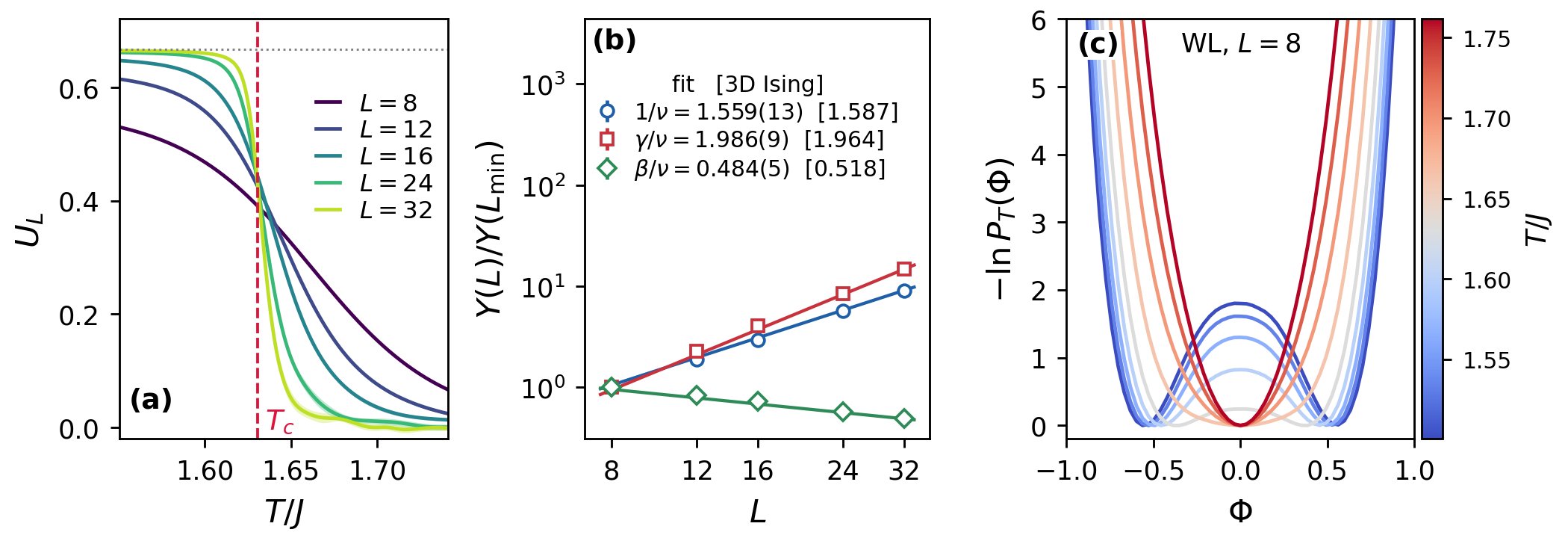}
	\caption{The star-order transition. (a) Binder cumulants $U_L$ from PT for $L=8$--$32$ (bands: jackknife errors); they cross at $T_c=1.6309(15)J$ (dashed line); dotted: ordered value $2/3$. (b) Finite-size scaling of the maximal slope of the Binder cumulant ($\propto L^{1/\nu}$, circles), of $\chi'(T_c)=L^3\langle\Phi^2\rangle/T_c$ ($\propto L^{\gamma/\nu}$, squares) and of $\langle|\Phi|\rangle(T_c)$ ($\propto L^{-\beta/\nu}$, diamonds), each normalized to its value at $L=8$; lines are power-law fits, with the exponents in the legend and the 3D Ising values in brackets. (c) Free-energy landscape $-\ln P_T(\Phi)$ from the WL joint histogram $H(E,\Phi)$ ($L=8$) at temperatures across the transition (color bar): the two minima at $\pm\Phi_0$ merge continuously at $\Phi=0$, without a coexisting disordered minimum.}
	\label{fig:fss}
\end{figure*}

Our residual entropy is closely related to that of the three-state (three-color) Potts antiferromagnet on the cubic lattice. Through $\phi\bmod3$, the ground states are the proper 3-colorings of the cubic lattice, i.e., the ground states of that model, and our $S_0$ per vertex, $3\times0.1223=0.367$, agrees with the value $0.3670(1)$ obtained there~\cite{YamaguchiOkabe2001,WangSwendsenKotecky1989}. Moreover, its low-temperature phase breaks the sublattice symmetry~\cite{WangSwendsenKotecky1989}, assigning one color to one sublattice and letting the other two fluctuate on the other, and the corresponding states give the lower bound $\frac12\ln2$ per vertex, i.e., our $\ln(2)/6$ per spin. The star order is the height-model counterpart of this phase. The two transitions differ, however: in the Potts model the ordered phase also selects a color, giving six ordered states and a transition in the 3D XY class~\cite{YamaguchiOkabe2001}, whereas here, there is no genuine $\mathbb Z_3$ symmetry, a global shift of $\phi$ leaves the spins unchanged, and only the $\mathbb Z_2$ sublattice symmetry is broken.

\emph{A 3D Ising transition.}---To establish the nature of the transition, we recorded $\Phi$ during the final WL stage, which yields unbiased microcanonical averages and the joint histogram $H(E,\Phi)$, hence the free-energy landscape $-\ln P_T(\Phi)$ at any temperature~\cite{Guruciaga2016}. The two minima at $\pm\Phi_0$ approach continuously and merge at $\Phi=0$, without a coexisting disordered minimum [Fig.~\ref{fig:fss}(c)], and $P_T(E)$ has a single peak (SM, Fig.~\ref{fig:sm_transition}(a)), indicating that the transition is continuous.

To determine its universality class, we used PT~\cite{HukushimaNemoto1996} with temperatures placed from the specific heat and multiple-histogram reweighting~\cite{FerrenbergSwendsen1989} (SM, Sec.~\ref{sm:methods}). The Binder cumulants $U=1-\langle\Phi^4\rangle/3\langle\Phi^2\rangle^2$ of $L=24$ and $32$ cross at $T_c=1.6309(15)J$ [Fig.~\ref{fig:fss}(a)]. Fits over $L=8$--$32$ [Fig.~\ref{fig:fss}(b) and Table~\ref{tab:exponents}] give $1/\nu=1.559(13)$, $\gamma/\nu=1.986(9)$ from $\chi'(T_c)=L^3\langle\Phi^2\rangle/T_c$ and $1.883(14)$ from the maximum of $\chi=L^3(\langle\Phi^2\rangle-\langle|\Phi|\rangle^2)/T$, and $\beta/\nu=0.484(5)$.  These values are close to the 3D Ising ones~\cite{Hasenbusch2010} and approximately satisfy hyperscaling, $2\beta/\nu+\gamma/\nu=2.95(1)$ versus $d=3$; we attribute the remaining deviations to corrections to scaling. The specific-heat maximum is described by $a+bL^{\alpha/\nu}$ with the Ising $\alpha/\nu=0.174$ and a negative background, as is common in this class. The main open points are $\beta/\nu$ and the crossing value $U^*\simeq0.38$, below the Ising $0.4655$ (SM, Sec.~\ref{sm:fss}). A 3D Ising transition is the natural expectation for the $\mathbb Z_2$ sublattice symmetry if the effective interactions between the charges $|Q|$ are short-ranged.

\emph{Discussion.}---Amp\`ere spin liquids are dual to Coulomb ones in two dimensions, and in three dimensions they carry a two-form Gauss law, three coupled two-dimensional Gauss laws, that, in principle, produce pinch points. What distinguishes them from Coulomb phases is that their charges are free. We have shown that in the octochlore ASL these free charges order within the ground-state manifold: an entropic sublattice order of the stars, invisible to Bragg scattering, that washes out the pinch-point correlations and sets in through a transition consistent with the 3D Ising class. Thus, entropic selection turns the system at low temperatures into an effective time-reversal invariant short-range spin liquid coexisting with a broken lattice symmetry. 
This scenario is reminiscent of fragmentation~\cite{lhotel2020fragmentation}, but in the magnitude of the charges rather than in their signs: there is no magnetic order, and $\mathbf m$ remains a pure gradient. Its signature for neutron scattering, or for real-space imaging in artificial realizations, is a set of strong, size-independent diffuse maxima at $\mathbf Q_0+\mathbf G$. 
Star-like configurations (all-in/all-out tetrahedra) also exist in the pyrochlore ASL~\cite{chung20252,rougemaille2025ampere} (where the spins sit on the links of a diamond lattice), but there they coexist with sharp, Bragg-like peaks~\cite{chung20252}, whereas here the star order gives only size-independent diffuse maxima; why the two bipartite lattices, diamond and cubic, behave differently is an open question. Away from the Amp\`ere point, a ratio $\alpha\neq1$ between the second- and first-neighbors couplings lifts the extensive degeneracy and selects layered ($\alpha>1$) states or all-in/all-out ($\alpha<1$), while the star order survives only in a temperature window close to $\alpha=1$ (SM, Sec.~\ref{sm:alpha}). 
The octochlore lattice was recently proposed as a platform for three-dimensional frustrated magnetism, with spin-ice and fracton liquids among its phases~\cite{stern2026fracton}.  Finally, the Amp\`ere point requires bond-dependent anisotropic exchange (SM, Sec.~\ref{sm:exchange}), generic for rare-earth ions with strong spin-orbit coupling~\cite{rau2019frustrated}; since the octochlore lattice is realized in rare-earth anti-perovskites and fluorides~\cite{stern2026fracton}, these materials are natural candidates in which to search for this regime.

\begin{acknowledgments}
\section*{Acknowledgments} 

It is a pleasure to thank Peter C. W. Holdsworth, Fr\'ed\'eric  Mila and  Na{\"\i}mo Davier for very enlightening discussions. We also thank the CNRS-funded IRP COQSYS for financial support that made this collaboration possible. HDR and FAGA are partially supported by CONICET (PIP 2021-112200200101480CO),  SECyT UNLP PI+D  X947 and X1065, and PICT-2020 - SERIEA-03205.
\end{acknowledgments}


%

\clearpage
\appendix
\counterwithout{equation}{section}
\setcounter{secnumdepth}{1} 

\setcounter{figure}{0}
\setcounter{table}{0}
\setcounter{equation}{0}
\renewcommand{\thefigure}{S\arabic{figure}}
\renewcommand{\thetable}{S\arabic{table}}
\renewcommand{\theequation}{S\arabic{equation}}
\renewcommand{\thesection}{S\arabic{section}}

\begin{center}
\textbf{Supplemental Material}
\end{center}

\section{Two dimensions: equivalence with the checkerboard Coulomb liquid}
\label{sm:2D}

\begin{figure}[h]
	\centering
	\includegraphics[width=\linewidth]{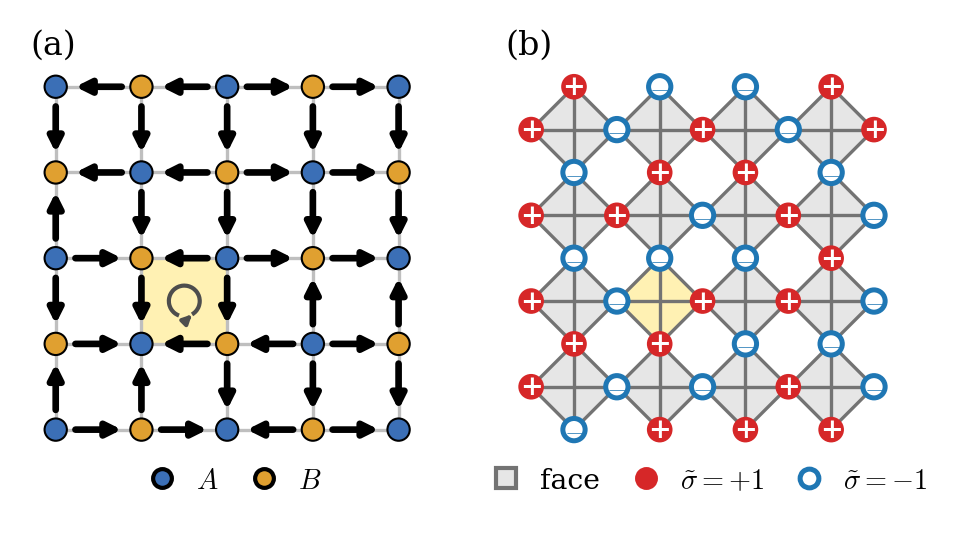}
	\caption{Equivalence between the square-lattice Amp\`ere model and the checkerboard antiferromagnet. (a) Arrows on the links of a ground state ($A$ blue, $B$ orange); every plaquette, e.g.\ the shaded one, has zero circulation. (b) Spins $\tilde\sigma=\epsilon\sigma$ on the checkerboard lattice (red: $+1$, blue: $-1$); crossed plaquettes (gray) are the original faces, where $\sum\tilde\sigma=0$; the shaded one is the face shaded in (a).}
	\label{fig:sm_2D}
\end{figure}

On the square lattice, with arrows on the links and the $A\to B$ convention, two parallel links of a plaquette carry opposite arrows for the same $\sigma$, and the circulation reads, up to an overall sign,
\begin{equation}
	\sigma_p=\sum_{\langle ij\rangle\in p}\epsilon_{ij}\,\sigma_{ij},\qquad
	\epsilon_{ij}=\begin{cases}+1&\text{vertical link},\\-1&\text{horizontal link},\end{cases}
\end{equation}
and $\mathcal{H}=\frac J2\sum_p\sigma_p^2$~\cite{rougemaille2025ampere}. With $\tilde\sigma_{ij}=\epsilon_{ij}\sigma_{ij}$, $\mathcal{H}=\frac J2\sum_p(\sum_{i\in p}\tilde\sigma_i)^2$. Placing $\tilde\sigma$ at the midpoints of the links, which form a square lattice rotated by $45^\circ$, the original plaquettes become the crossed plaquettes of a checkerboard lattice and the original vertices its blank plaquettes; the Hamiltonian is the antiferromagnetic Ising model on the checkerboard lattice, a CSL~\cite{davier2023combined} (Fig.~\ref{fig:sm_2D}). The two models have the same ground states and the same correlations of $\tilde\sigma$. The variables $\tilde\sigma$ are, up to a sign, the rotated arrows of Eq.~(\ref{eq:E2D}): on the dual lattice of square centers (sublattices $a$, $b$), for $\sigma=+1$ the dual arrow crossing a horizontal link points from $b$ to $a$ and the one crossing a vertical link from $a$ to $b$, and the net flux at a dual site is $-\Gamma_p$. The exact mapping does not extend to three dimensions, where a face has four spins and an octahedron six.
%


\section{Exchange couplings of the Amp\`ere model}
\label{sm:exchange}

The moments are $\boldsymbol\mu_i=\sigma_i\mathbf d_i$, with $\mathbf d_i=\eta_i\hat e_\mu$ the easy axis along the link, oriented from $A$ to $B$, as in Ref.~\cite{stern2026fracton}. For Ising ions, a general bilinear exchange $\boldsymbol\mu_i\cdot\mathcal J_{ij}\cdot\boldsymbol\mu_j$ reduces to $J^{\rm eff}_{ij}\sigma_i\sigma_j$, with
\begin{equation}
J^{\rm eff}_{ij}=\mathbf d_i\cdot\mathcal J_{ij}\cdot\mathbf d_j .
\end{equation}
Expanding Eq.~(\ref{eq:H}) gives $J^{\rm eff}=-J$ for the NN$_1$ pairs (distance $a/\sqrt2$), $+J$ for the parallel pairs across a face ($J_{2b}$, distance $a$) and $0$ for the collinear pairs of an octahedron ($J_{2a}$, distance $a$). Isotropic exchange, $\mathcal J_{ij}\propto\openone$, gives $J^{\rm eff}\propto\mathbf d_i\cdot\mathbf d_j=0$ for NN$_1$ pairs, whose axes are perpendicular, so a nonzero $J_1$ requires anisotropic exchange, here as in Ref.~\cite{stern2026fracton}. The three values follow from the transverse form
\begin{equation}
\mathcal J_{ij}=c_{ij}\,(\openone-\hat{\mathbf r}_{ij}\hat{\mathbf r}_{ij}),
\end{equation}
which couples only the moment components perpendicular to the bond. For $J_{2a}$ pairs both axes lie along the bond, so $J^{\rm eff}_{2a}=0$ identically. For NN$_1$ pairs, $\mathbf d_i\cdot\mathbf d_j=0$ and $(\mathbf d_i\cdot\hat{\mathbf r})(\mathbf d_j\cdot\hat{\mathbf r})=-\frac12$, so $J^{\rm eff}_1=c_1/2$; for $J_{2b}$ pairs, $\mathbf d_i\cdot\mathbf d_j=-1$ and $\mathbf d_{i,j}\perp\hat{\mathbf r}$, so $J^{\rm eff}_{2b}=-c_{2b}$. The Amp\`ere model is $c_1=-2J$, $c_{2b}=-J$, i.e., $c_{ij}=-Ja^2/r_{ij}^2$ for $r_{ij}\le a$: a ferromagnetic coupling of the transverse components of the moments. 

\section{Rank-two Gauss law on the octochlore lattice}
\label{sm:gauss}

\begin{figure*}[t]
	\centering
	\includegraphics[width=0.8\textwidth]{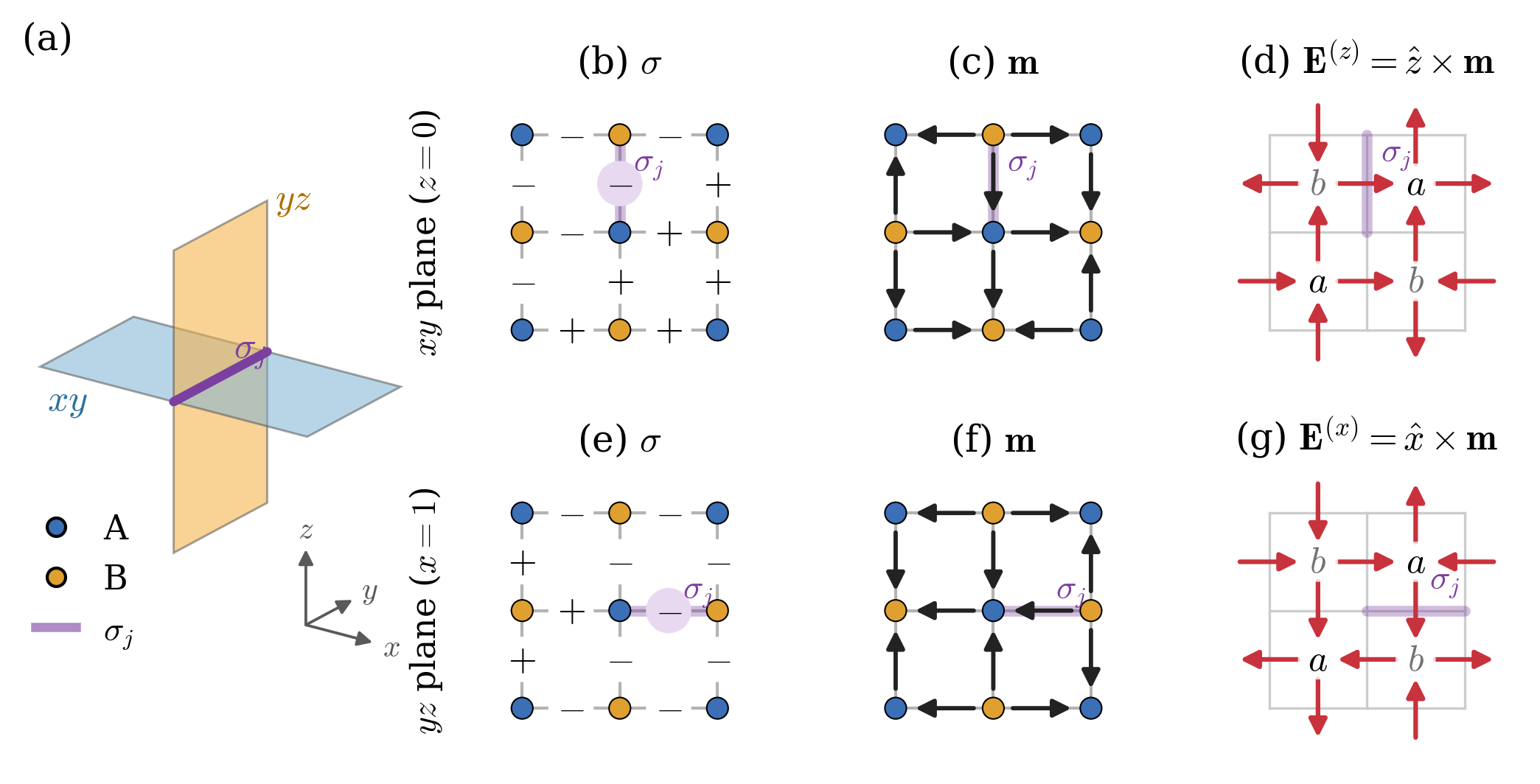}
	\caption{Construction of the Gauss law for a ground state of the Ising model. (a) A spin $\sigma_j$ on a $y$ link belongs to two $xy$ and two $yz$ faces. (b,e) Ising variables in the $xy$ plane $z=0$ and in the $yz$ plane $x=1$ (vertices of sublattice $A$ blue, $B$ orange). (c,f) Arrows $\mathbf m=\eta\sigma$, from $A$ to $B$ for $\sigma=+1$. (d,g) Rotated arrows $\mathbf E^{(z)}=\hat z\times\mathbf m$ and $\mathbf E^{(x)}=\hat x\times\mathbf m$ (columns $z$ and $x$ of $E_{\mu\nu}$) on the dual lattice of face centers $a$, $b$, with zero net flux at every face center. The same spin gives two different arrows, $E_{xz}=-m_y$ in (d) and $E_{zx}=+m_y$ in (g).}
	\label{fig:sm_gauss}
\end{figure*}

Label a link by its lower vertex $\mathbf r$ and direction $\mu$, so that $m_{\mathbf r,\mu}=\eta_{\mathbf r}\sigma_{\mathbf r,\mu}$ and the circulation around a face with lower corner $\mathbf r$ is $\eta_{\mathbf r}\mathcal S_p$. Each link belongs to four faces, two in each of the two families of planes that contain it [Fig.~\ref{fig:sm_gauss}(a)]. In each family of normal $\hat e_\nu$ we place dual sites at the face centers, split into $a$ and $b$ according to the sublattice of the lower corner, and draw on the dual links the rotated arrows $\mathbf E^{(\nu)}=\hat e_\nu\times\mathbf m$. In terms of the Ising variables, for $\sigma=+1$ the dual arrow crossing a link parallel to $\hat u$ goes from $b$ to $a$, and the one crossing a link parallel to $\hat v$ goes from $a$ to $b$, with $(u,v,\nu)$ in cyclic order. Each spin produces two different dual arrows,
\begin{equation}
	\begin{array}{lll}
	\text{link along } x: & E^{(z)}_y=+m_x, & E^{(y)}_z=-m_x,\\[1pt]
	\text{link along } y: & E^{(x)}_z=+m_y, & E^{(z)}_x=-m_y,\\[1pt]
	\text{link along } z: & E^{(y)}_x=+m_z, & E^{(x)}_y=-m_z ,
	\end{array}
\end{equation}
which are the entries $E_{\mu\nu}=E^{(\nu)}_\mu$ of the antisymmetric tensor $E_{\mu\nu}=\epsilon_{\mu\nu\rho}m_\rho$ of Eq.~(\ref{eq:gauss}). Explicitly,
\begin{equation}
	E=\begin{pmatrix} 0 & m_z & -m_y\\ -m_z & 0 & m_x\\ m_y & -m_x & 0\end{pmatrix},
\end{equation}
whose column $\nu$ is $\mathbf E^{(\nu)}=\hat e_\nu\times\mathbf m$, a two-dimensional vector in the planes perpendicular to $\hat e_\nu$. The Amp\`ere law on a face of normal $\hat e_\nu$ is the two-dimensional Gauss law of this column at the face center [Fig.~\ref{fig:sm_gauss}(d,g)], e.g.\ $\partial_xE_{xz}+\partial_yE_{yz}=-\partial_xm_y+\partial_ym_x=0$ for the $xy$ faces. The three laws are coupled because the same $m_\rho$ appears in two columns, and they are linearly dependent, $\partial_\nu\partial_\mu E_{\mu\nu}\equiv0$ by antisymmetry, which on the lattice is the vanishing of the sum of the six circulations of a cube.

The models are summarized in Table~\ref{tab:sm_summary}.
\begin{table}[h]
\caption{Ising Amp\`ere models and their Gauss laws.}
\label{tab:sm_summary}
\begin{ruledtabular}
\begin{tabular}{lll}
Model & Ground states & Gauss law \\
\hline
Square & heights $\phi\in\mathbb Z$ & $\partial_\mu E_\mu=0$, $E_\mu=\epsilon_{\mu\nu}m_\nu$ \\
Octochlore & heights $\phi\in\mathbb Z$ & $\partial_\mu E_{\mu\nu}=0$, $E_{\mu\nu}=\epsilon_{\mu\nu\rho}m_\rho$ \\
Pyrochlore & heights on diamond & as octochlore \\
\end{tabular}
\end{ruledtabular}
\end{table}

\section{Constraint vectors and structure factor}
\label{sm:constraint}
 
We analyze the correlations with the constraint-vector method \cite{henley2005power,benton2021topological,davier2023combined}. Let $m_\mu(\mathbf q)=\sum_{\mathbf r}e^{-i\mathbf q\cdot(\mathbf r+\hat e_\mu/2)}m_{(\mathbf r,\mu)}$ be the Fourier transform of the arrow field of sublattice $\mu$. The circulation around a face of normal $\hat e_\rho$ is a linear combination of the four $m_\mu(\mathbf q)$ of its links, $\Gamma_\rho(\mathbf q)=\mathbf L_\rho(\mathbf q)\cdot\mathbf m(\mathbf q)$, with
\begin{equation}\label{eq:Lvec}
	L_{\rho,\mu}(\mathbf q)=\sum_{i\in p,\;i\in\mu}\tau_{ip}\,e^{i\mathbf q\cdot(\mathbf r_c-\mathbf r_i)},
\end{equation}
where $\mathbf r_c$ is the face center, $\mathbf r_i$ the position of the spin, and $\tau_{ip}=+1$ ($-1$) if the direction $\hat e_\mu$ of link $i$ coincides with (is opposite to) the counterclockwise sense of the face $p$ seen from $+\hat e_\rho$. One finds $	(\mathbf L_\rho)_\mu=2i\,\epsilon_{\rho\mu\nu}\,s_\nu$, with 
\begin{eqnarray}
\label{eq:Lexplicit}
	\mathbf L_x&=&2i\,(0,\,s_z,\,-s_y),\nonumber\\
	\mathbf L_y&=&2i\,(-s_z,\,0,\,s_x),\\
	\mathbf L_z&=&2i\,(s_y,\,-s_x,\,0),\nonumber
\end{eqnarray}
with $s_\mu=\sin(q_\mu/2)$. For instance, on the $xy$ face the two links along $x$ sit at $\mathbf r_c\pm\frac12\hat y$ and the two along $y$ at $\mathbf r_c\pm\frac12\hat x$, so that $L_{z,x}=e^{iq_y/2}-e^{-iq_y/2}=2is_y$ and $L_{z,y}=e^{-iq_x/2}-e^{iq_x/2}=-2is_x$. The three vectors are just the components of a lattice curl, $\mathbf L_\rho\!\cdot\!\mathbf m=2i\,(\mathbf m\times\mathbf s)_\rho$, with $\mathbf s=(s_x,s_y,s_z)$. They obey $\sum_\rho s_\rho\mathbf L_\rho=0$, which is the Fourier-space form of the relation between the face circulations of a cube [and of $\partial_\nu\partial_\mu E_{\mu\nu}\equiv0$].  Therefore the matrix
\begin{equation}
	\mathbf M(\mathbf q)=\bigl[\mathbf L_x(\mathbf q),\mathbf L_y(\mathbf q),\mathbf L_z(\mathbf q)\bigr]
\end{equation}
has rank two (with the $\mathbf L_\rho$ as columns, $\mathbf M$ is singular and the inverse in the projector must be a pseudo-inverse, denoted by $+$; equivalently one may keep two independent vectors, as in Ref.~\cite{davier2023combined}). The constraints $\mathbf L_\rho\!\cdot\!\mathbf m=0$ force $\mathbf m(\mathbf q)\parallel\mathbf s$, i.e. $m_\mu(\mathbf q)=2is_\mu\phi(\mathbf q)$, which is the Fourier transform of the gradient $m_{(\mathbf r,\mu)}=\phi(\mathbf r+\hat e_\mu)-\phi(\mathbf r)$. The correlation function is the projector on the space allowed by the constraints \cite{henley2005power}, Eq.~(\ref{eq:projector}), $\big\langle m_\mu(-\mathbf q)\,m_\nu(\mathbf q)\big\rangle\propto \Pi_{\mu\nu}(\mathbf q)$ with $\Pi=1-\mathbf M\big(\mathbf M^\dagger\mathbf M\big)^{+}\mathbf M^\dagger=\frac{\mathbf s\,\mathbf s^{T}}{|\mathbf s|^2}$. 
For a single constraint vector it reduces to the usual $\Pi=1-\mathbf L\mathbf L^\dagger/\|\mathbf L\|^2$~\cite{davier2023combined}.
It is exact for the Gaussian (soft-constraint) version of the model and is an approximation for hard spins; the Monte Carlo results test it. The points $\mathbf q=2\pi\mathbf n$ ($\mathbf s=0$), where all constraints are trivially satisfied, are singular and are left blank in the figures.

\emph{Structure factor.} We use the same definition as for spin ice~\cite{Isakov_Dipolar,fennell2009magnetic}. The moments are along the links, $\boldsymbol\mu_i=\sigma_i\,\mathbf d_i$, where $\mathbf d_i=\eta_i\hat e_\mu$ is the easy axis oriented from $A$ to $B$ (as the local $\langle111\rangle$ axes of spin ice), so that the moment is the arrow $\mathbf m_i$. The structure factor of Eq.~(\ref{eq:Sq_def}), normalized per spin ($N=3L^3$), is computed in MC from $\mathbf m_\mu(\mathbf q)=\sum_{i\in\mu}m_i\,e^{-i\mathbf q\cdot\mathbf r_i}$ as $S(\mathbf q)=\frac1N\sum_{\mu\nu}(\delta_{\mu\nu}-\hat q_\mu\hat q_\nu)\langle m_\mu(-\mathbf q)m_\nu(\mathbf q)\rangle$, averaged over the stored configurations. With $\langle m_\mu(-\mathbf q)m_\nu(\mathbf q)\rangle/N=\Pi_{\mu\nu}(\mathbf q)$, the projector gives
\begin{equation}\label{eq:Sq_ising}
	\mathbb S(\mathbf q)=\sum_{\mu\nu}\Big(\delta_{\mu\nu}-\frac{q_\mu q_\nu}{q^2}\Big)\Pi_{\mu\nu}(\mathbf q)=1-\frac{(\mathbf q\cdot\mathbf s)^2}{q^2\,|\mathbf s|^2},
\end{equation}
i.e., Eq.~(\ref{eq:Sq_def}) with the correlations of the components perpendicular to $\mathbf q$ given by the projector. It has direction-dependent singularities at $\mathbf q=2\pi\mathbf n$ (pinch points). The polarization factor is essential here: since the three easy axes are orthogonal, the unprojected trace $\sum_\mu\langle m_\mu(-\mathbf q)m_\mu(\mathbf q)\rangle$ is featureless in the projector approximation ($\mathrm{Tr}\,\Pi=1$), and the pinch points appear only through the distribution of the intensity among the components. $\mathbb S(\mathbf q)$ is bounded by one, which the maxima observed at $\mathbf Q_0+\mathbf G$ exceed. The correlations of the raw variables $\sigma_i$ (without the factor $\eta_i$) are those of the arrow field shifted by $\mathbf Q_0=(\pi,\pi,\pi)$. 
\section{Methods}
\label{sm:methods}

\emph{Annealing MC.} Independent copies start from random configurations and are cooled as $T_n=T_0r^n$ ($T_0=10J$, $r=0.95$, $n=1,\dots,70$, i.e., from $9.5J$ down to $0.276J$; $10^4$ equilibration sweeps and $10^4$ measurements every 5 sweeps per temperature; 100 copies; 5 stored configurations per copy and temperature). A sweep of single-spin Metropolis flips is followed by a star move: a random vertex is chosen and, if its six spins are equal, they are all reversed. This costs no energy (the two links of a face meeting at the vertex enter $\mathcal S_p$ with opposite signs) and shifts $\phi$ by $\pm2$. Global winding sectors cannot be changed by local updates.

\emph{Wang-Landau.} With $E=\frac J2\sum_p\mathcal S_p^2$, energies are multiples of $4J$, and one bin per level is used up to six standard deviations above the maximum of $g(E)$ ($\langle E\rangle_\infty=6JL^3$, $\mathrm{Var}_\infty=18J^2L^3$), which omits $\lesssim10^{-9}$ of the states. Single flips are accepted with $\min[1,g(E)/g(E')]$; star moves leave $E$ unchanged and are always accepted~\cite{WangLandau2001}. We use an $80\%$ flatness criterion followed by the $1/t$ scheme~\cite{BelardinelliPereyra2007} down to $\ln f=10^{-8}$, with $8$ independent runs per size (errors from their dispersion), and normalize with $\sum_Eg(E)=2^N$. For $\ln f\le10^{-5}$ we accumulate $\langle|\Phi|\rangle_E$, $\langle\Phi^2\rangle_E$, $\langle\Phi^4\rangle_E$ and the joint histogram $H(E,\Phi)$, which are unbiased because the weights depend on $E$ only; canonical averages follow as $\langle A\rangle_T=\sum_Eg(E)e^{-E/T}\langle A\rangle_E/Z$ and $P_T(\Phi)\propto\sum_Eg(E)e^{-E/T}H(E,\Phi)/H(E)$~\cite{Guruciaga2016}. The code was validated against the exact enumeration of the $2^{24}$ states of $L=2$, performed with the Hamiltonian of the MC code: $\max|\Delta\ln g|=7\times10^{-4}$, $S_0/N=0.19428(2)$ versus $0.194310$, and exact microcanonical $\langle\Phi^2\rangle_E$.

\begin{figure}[t]
	\centering
	\includegraphics[width=0.8\linewidth]{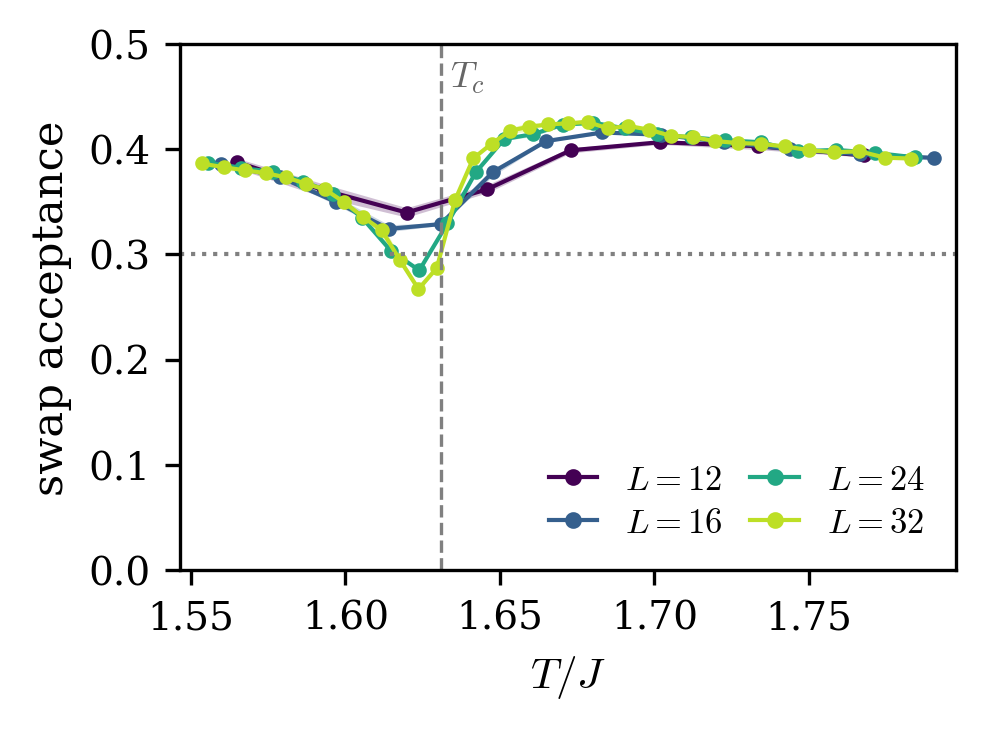}
	\caption{PT diagnostics: exchange acceptance between neighbouring temperatures $T_i$, $T_{i+1}$, plotted at their midpoint, for $L=12$--$32$ (same window $1.55$--$1.80J$); average over runs (symbols) and range between runs (bands). Dotted: target acceptance $0.3$; dashed: $T_c$. The acceptance has its minimum at $T_c$ and stays above $0.26$. For $L=8$ (uniform grid of $16$ temperatures in $1.50$--$1.85J$) it is above $0.67$.}
	\label{fig:sm_PT_diag}
\end{figure}

\emph{Parallel tempering.} One replica per temperature evolves with single flips and star moves; exchanges between neighbouring temperatures are attempted after every sweep, alternating even and odd pairs~\cite{HukushimaNemoto1996}. The temperatures cover $1.50$--$1.85J$ ($L=8$) and $1.55$--$1.80J$ ($L\ge12$), with $16$, $9$, $14$, $24$ and $36$ replicas for $L=8$, $12$, $16$, $24$, $32$; for $L\ge12$ they are spaced as $\Delta\beta=2\,\mathrm{erfc}^{-1}(0.3)/\sigma_E(T)$ with $\sigma_E^2=T^2C_v$ from a preliminary run, which targets a uniform acceptance $\simeq0.3$. We performed $2\times10^4$ thermalization and $10^5$ measurement sweeps, recording $E$ and the star numbers $n_A$, $n_B$ every two sweeps, with $4$ runs for $L\le16$ and $2$ for $L=24$, $32$. For $L=8$ the grid is uniform. The acceptance has its minimum at $T_c$ and stays above $0.26$ (above $0.67$ for $L=8$; Fig.~\ref{fig:sm_PT_diag}), and each run performs $\sim10^3$ round trips between the extreme temperatures. All time series are combined with the multiple-histogram method in the energy~\cite{FerrenbergSwendsen1989}, with microcanonical averages of the observables at each energy, and errors are obtained by jackknife over 10 blocks. For $L=8$ PT and WL coincide. Additional parallel-tempering simulations for $L=40$ were used for the structure-factor analysis shown in Fig.~\ref{sm:Sq}; this size was not included in the critical finite-size scaling analysis.
%

\section{Thermodynamics: details}
\label{sm:thermo}

\begin{figure}[t]
	\centering
	\includegraphics[width=0.99\linewidth]{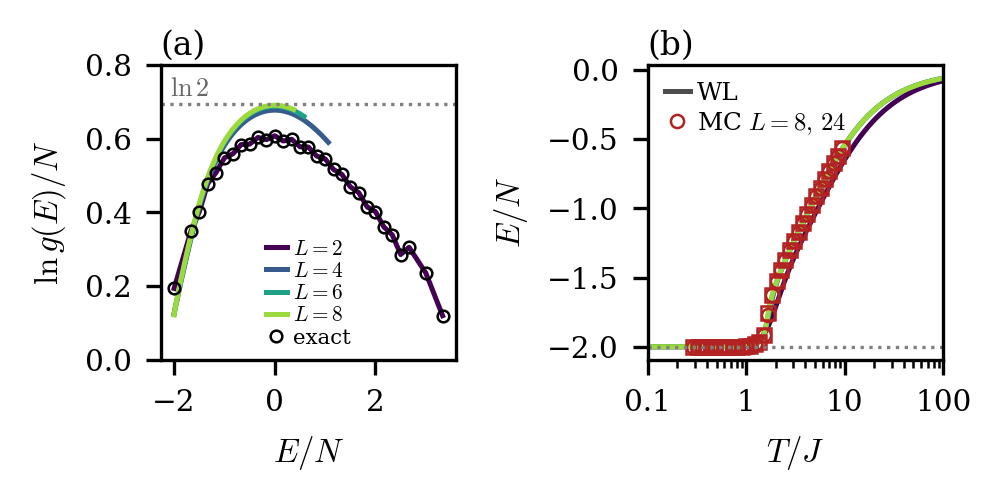}
	\caption{(a) $\ln g(E)/N$ from WL ($L=2$--$8$) and exact enumeration ($L=2$, circles), normalized to $\sum_Eg=2^N$; dotted: $\ln2$. (b) Energy per spin from WL (lines) and annealing MC ($L=8$, $24$; symbols), with $E/N=-2J$ in the ground states (MC convention).}
	\label{fig:sm_lng}
\end{figure}

\emph{Density of states and entropy.} Figure~\ref{fig:sm_lng} shows $\ln g(E)/N$ for $L=2$--$8$ and the energy per spin, which agrees with the annealing MC. The entropy of Fig.~\ref{fig:thermo}(a) follows from $g(E)$ as $S(T)=\ln Z+\langle E\rangle/T$, with the absolute normalization fixed by $\sum_Eg(E)=2^N$, i.e., by the paramagnetic limit $S/N\to\ln2$. A fit $S_0(L)=S_0^\infty+a/L^3$ for $L\ge4$ (with periodic boundaries the corrections are $O(1/N)$) gives $S_0^\infty/N=0.1223(1)$, $2.7\%$ above the exact bound $\frac{33}{32}\frac{\ln2}{6}\simeq0.1191$ of Eq.~(\ref{eq:bound}). Through $\phi\bmod3$ the ground states are the proper 3-colorings of the cubic lattice~\cite{WangSwendsenKotecky1989}; on the torus the winding sectors of heights and colorings differ, which only affects subextensive terms.  Per vertex, $3S_0/N=0.3669(3)$ agrees with the ground-state entropy $0.3670(1)$ of the three-state Potts antiferromagnet on the cubic lattice~\cite{YamaguchiOkabe2001}.

\emph{Specific heat from three methods.} $C_v$ was obtained from the energy fluctuations of the annealing MC (100 copies), from the WL density of states, and from PT with multiple-histogram reweighting. All agree where they overlap [Fig.~\ref{fig:thermo}(b)]. The geometric grid of the annealing ($T_{n+1}/T_n=0.95$) is too coarse near $T_c$, so its maximum for $L=24$ ($0.96$ at $1.66J$) underestimates the PT one ($1.77$). At high temperature all approach the exact limit $C_v/N=6J^2/T^2$. The maxima are $C_v^{\max}/N=0.87$, $1.05$, $1.18$ at $T/J=1.692$, $1.654$, $1.638$ for $L=4$, $6$, $8$ (WL), and $1.18$, $1.38$, $1.51$, $1.77$, $1.90$ for $L=8$--$32$ (PT); their size dependence is discussed in Sec.~\ref{sm:fss}.

\section{Star order and nature of the transition}
\label{sm:order}

\emph{Why the plateaus are not $1$ and $1/32$.} A vertex is a star when its six neighbours have the same height. In the bound family ($\phi=0$ on $A$, $\phi=\pm1$ at random on $B$) all $B$ vertices are stars and an $A$ vertex is one with probability $2/2^6=1/32$; this family has entropy $\ln(2)/6$ and is an exponentially small part of the ground states. Since all ground states have the same energy, the plateaus are those of the most numerous ground states, which differ from this family in two ways. First, an $A$ star can be shifted by $\pm2$ at no energy cost (a star move): it remains a star, but its six $B$ neighbours stop being stars. Each $A$ star thus doubles the number of ground states with a given sign configuration of $B$, and in typical states half of them are shifted; with random signs on $B$ this alone gives $f_B\simeq(1-1/64)^6\simeq0.91$ and the bound in Eq.~(\ref{eq:bound}). Second, sign configurations of $B$ with more $A$ stars carry more ground states, which favours local alignment of the $B$ signs: two $A$ vertices at distance $\sqrt2$ share two $B$ neighbours, so the probability that both are stars is $2^{-9}$, twice the product of the single probabilities. Monte Carlo sampling of the restricted ground-state ensemble with $\phi\in\{0,\pm2\}$ on $A$ and $\phi=\pm1$ on $B$, i.e., the measure $\propto\lambda^{n_A(s)}$ on the signs of $B$ at $\lambda=2$ (MC, $L=16$, $20$), gives $f_A=0.125$ and $f_B=0.717$, and integrating $\langle n_A\rangle$ over $1\le\lambda\le2$ gives $S_0/N=\frac16[\ln2+\int_1^2f_A(\lambda)\,d\lambda/\lambda]=0.1223$, in agreement with the full simulations ($f_{\min}\simeq0.13$, $f_{\max}\simeq0.71$, $S_0/N=0.1223(1)$). Typical ground states are therefore the flat family dressed by these zero-energy rearrangements; larger excursions of the heights are negligible. The same ensemble reproduces the diffuse maxima of $S(\mathbf q)$: with the definition of Eq.~(\ref{eq:Sq_def}), it gives $S(\mathbf Q_p)=7.1(5)$ and $7.2(7)$ for $L=16$ and $24$, compared with $6.7(4)$ and $6.3(4)$ in the full MC at $T=0.276J$, versus $S=|\mathbf F_\perp|^2/6=1.94$ for uncorrelated signs ($\lambda=1$), with $\mathbf F=2i\mathbf s$ the form factor of a star. In the language of 3-colorings ($\phi\bmod3$) this is the statement that typical colorings of a bipartite lattice are nearly monochromatic on one sublattice, proven for high dimension~\cite{Peled2017} and observed numerically for $d=3$~\cite{WangSwendsenKotecky1989,YamaguchiOkabe2001}. Defects cost at least $8J$ (one spin flip changes four $\mathcal S_p$ by $\pm2$), hence $\rho_{\rm def}\propto e^{-8J/T}$ at low $T$.

\begin{figure*}[t]
	\centering
	\includegraphics[width=0.95\textwidth]{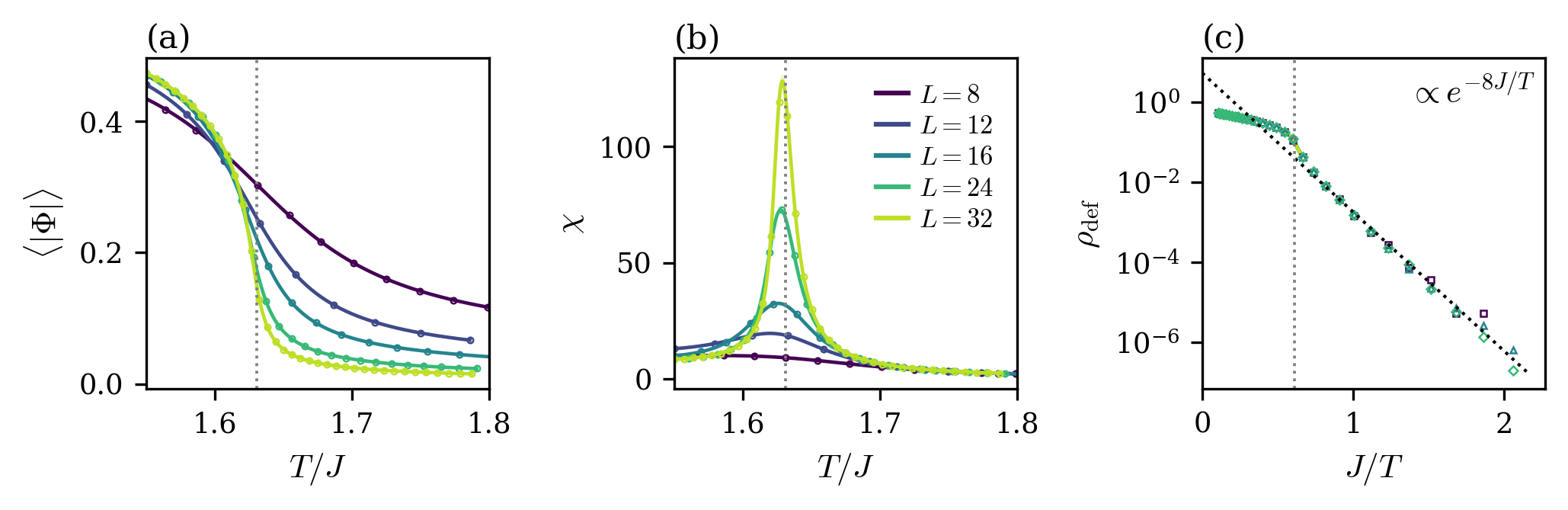}
	\caption{(a,b) PT results for $L=8$--$32$ in the common window $1.55$--$1.80J$: reweighted curves (lines, jackknife bands) and direct averages at the simulated temperatures (circles); dotted: $T_c$. (a) $\langle|\Phi|\rangle$; (b) $\chi=L^3(\langle\Phi^2\rangle-\langle|\Phi|\rangle^2)/T$. (c) Fraction $\rho_{\rm def}$ of faces with $\mathcal S_p\neq0$ versus $J/T$, from annealing MC over the whole range (symbols; zeros omitted) and PT near $T_c$ (lines); dotted: $e^{-8J/T}$, the Boltzmann factor of the smallest defect (one flipped spin, a closed string of four faces); vertical: $J/T_c$. $C_v$, the Binder cumulant and the star fractions are shown in Figs.~\ref{fig:thermo}(b), \ref{fig:fss}(a) and \ref{fig:order}(a).}
	\label{fig:sm_PT_obs}
\end{figure*}

The distribution of $\Phi$ from MC (Fig.~\ref{fig:order}) is centred at zero above $T_c$ and bimodal below, with peaks at $\pm\Phi_0$ that sharpen with $L$. The three methods agree on the ordered phase: $\langle|\Phi|\rangle\simeq0.55$ and $f\simeq0.69/0.14$ for $L=8$ from WL, MC and PT, and the PT curves for $L=8$--$32$ (Fig.~\ref{fig:sm_PT_obs}) show how the order parameter, its susceptibility, the star fractions and the density of defects evolve across the transition.

\begin{figure}[t]
	\centering
	\includegraphics[width=0.99\linewidth]{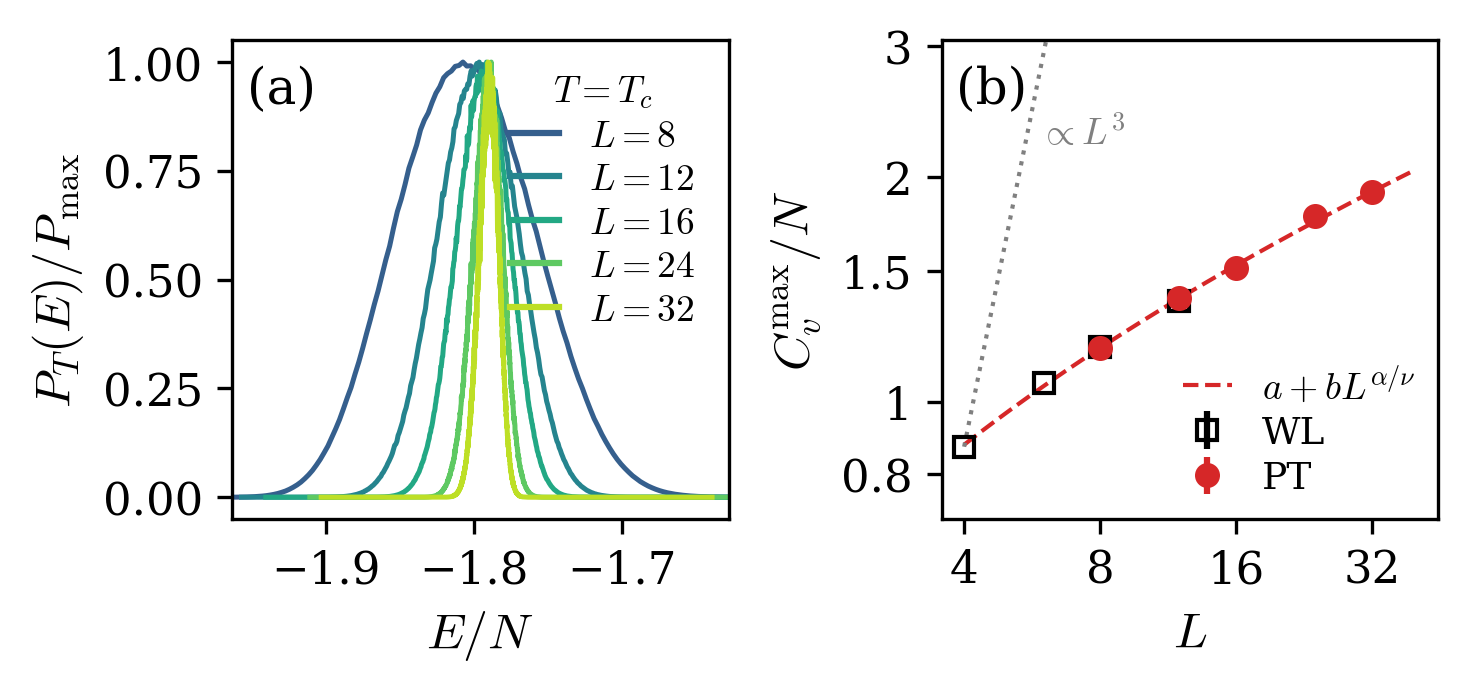}
	\caption{Nature of the transition. (a) Energy distribution $P_T(E)$ at $T_c=1.6309J$ from PT with multiple-histogram reweighting, $L=8$--$32$: a single peak that narrows with $L$. (b) $C_v^{\max}/N$ versus $L$ from WL (squares, $L=4$--$8$) and PT (circles, $L=8$--$32$); dashed: $a+bL^{\alpha/\nu}$ with the 3D Ising $\alpha/\nu=0.174$; dotted: the $L^3$ scaling of a first-order transition.}
	\label{fig:sm_transition}
\end{figure}

Three independent tests show that the transition is continuous. (i) The energy distribution at $T_c$, $P_T(E)\propto g(E)e^{-E/T}$, has a single peak for all sizes up to $L=32$ [Fig.~\ref{fig:sm_transition}(a)], whereas a first-order transition would give two. (ii) The landscape $-\ln P_{T}(\Phi)$ is a Landau free energy measured without approximation [Fig.~\ref{fig:fss}(c)]: for a first-order transition three minima ($0$ and $\pm\Phi_0$) would coexist at $T_c$, separated by barriers growing as $L^2$; here the two minima at $\pm\Phi_0$ merge continuously at $\Phi=0$, and at $T_c$ the barrier between them stays of order one ($0.25$--$0.6$ for $L=8$--$32$, from PT). (iii) $C_v^{\max}$ grows far more slowly than the $L^3$ of a first-order transition [Fig.~\ref{fig:sm_transition}(b)]; over $L=4$--$32$ it is described by the weak Ising divergence discussed in Sec.~\ref{sm:fss}, and WL and PT agree where they overlap ($L=8$).
%
\section{Critical exponents and finite-size scaling}
\label{sm:fss}

\emph{Wang-Landau, $L\le8$.} The Binder cumulants of $L=6$ and $8$ cross at $1.662J$ and the maximal slope gives $1/\nu=1.57$, but $\beta/\nu\simeq0.59$ and $\gamma/\nu\simeq1.49$ are far from any standard class and hyperscaling fails by $10\%$. These sizes are dominated by corrections; with PT the crossing drifts down to $1.631J$ and the exponents become Ising-like.

\emph{Parallel tempering, $L=8$--$32$.} The fits of Table~\ref{tab:exponents} use $\max|dU/dT|\propto L^{1/\nu}$, $\chi'(T_c)$ and $\chi_{\max}\propto L^{\gamma/\nu}$ ($\chi_{\max}$ for $L\ge12$; for $L=8$ $\chi$ has no maximum in the simulated range), $\langle|\Phi|\rangle(T_c)\propto L^{-\beta/\nu}$, and $C_v^{\max}$. Adding $L=32$ to the $L=8$--$24$ analysis moved $T_c$ from $1.6294(7)$ to $1.6309(15)$ and the exponents towards the Ising values ($\gamma/\nu$ from $\chi'$: $2.055\to1.986$; $\beta/\nu$: $0.440\to0.484$), as expected if the $T_c$ from smaller pairs is slightly low. A pure power law for $C_v^{\max}$ gives $0.363(5)$, but $C_v^{\max}/N=a+bL^{0.174}$ with $a\simeq-1.6$ and $b\simeq1.95$ describes the data (e.g.\ $1.39$ and $1.94$ predicted for $L=12$, $32$, versus $1.38$ and $1.90$). The low crossing value $U^*\simeq0.38$ (Ising: $0.4655$ for cubic systems with periodic boundaries~\cite{Hasenbusch2010}) is the main open point; it may reflect slow corrections associated with the composite nature of $\Phi$, since stars also exist in the disordered phase.

\section{Size and temperature dependence of the structure factor}
\label{sm:Sq}

\begin{figure*}[t]
	\centering
	\includegraphics[width=0.8\textwidth]{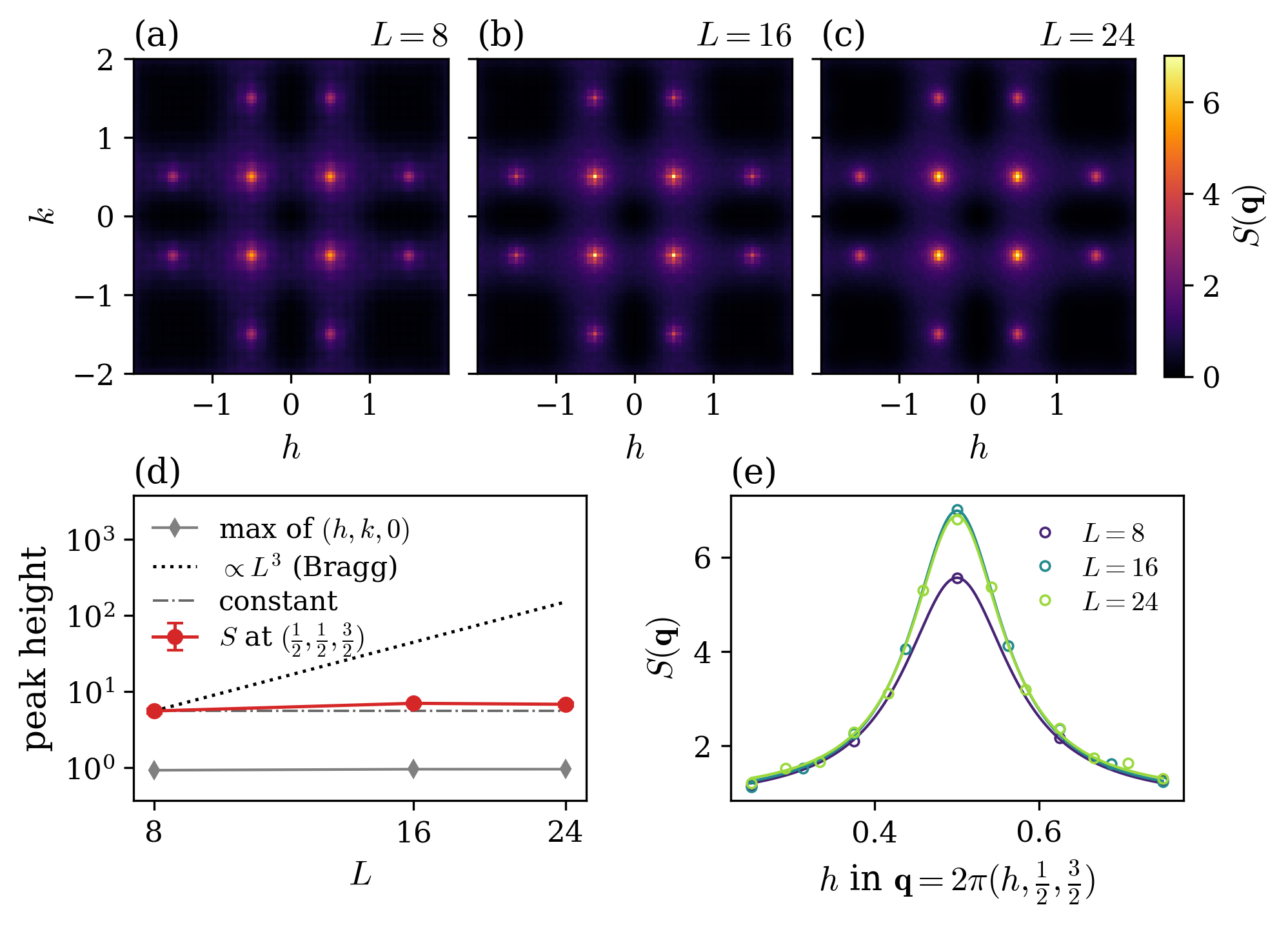}
	\caption{Size dependence of the maxima at $\mathbf Q_0+\mathbf G$, MC at $T\simeq0.28J$, 500 configurations per size. (a--c) $S(\mathbf q)$ in the $(h,k,\frac32)$ plane for $L=8$, $16$, $24$, common color scale. (d) $S$ at $\mathbf q=2\pi(\frac12,\frac12,\frac32)$ versus $L$ (log scale; errors from blocks of copies), with the maximum of the $(h,k,0)$ plane as a control, the Bragg scaling $L^3$ (dotted) and a constant (dash-dotted). (e) Cuts $S(h,\frac12,\frac32)$ at the allowed momenta $h=n/L$ (circles) with Lorentzian fits (lines).}
	\label{fig:sm_Sq_size}
\end{figure*}

\begin{figure}[t]
	\centering
	\includegraphics[width=0.99\linewidth]{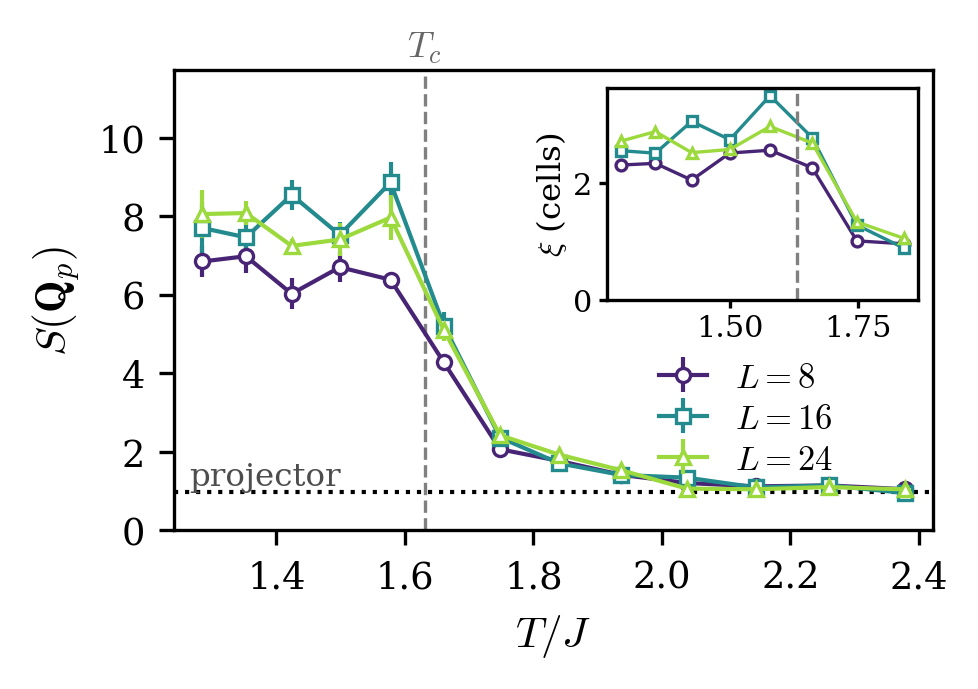}
	\caption{Maximum at $\mathbf Q_p=2\pi(\frac12,\frac12,\frac32)$ versus $T$ for $L=8$, $16$, $24$ (annealing MC, 500 configurations per point); dotted: projector value $0.97$; dashed: $T_c$. Inset: correlation length $\xi=1/(\pi\,\mathrm{FWHM})$, in cubic cells, from Lorentzian fits along $[100]$, $[010]$, $[001]$, shown where a peak exists.}
	\label{fig:sm_Sq_T}
\end{figure}

\begin{figure*}[t]
	\centering
	\includegraphics[width=0.85\textwidth]{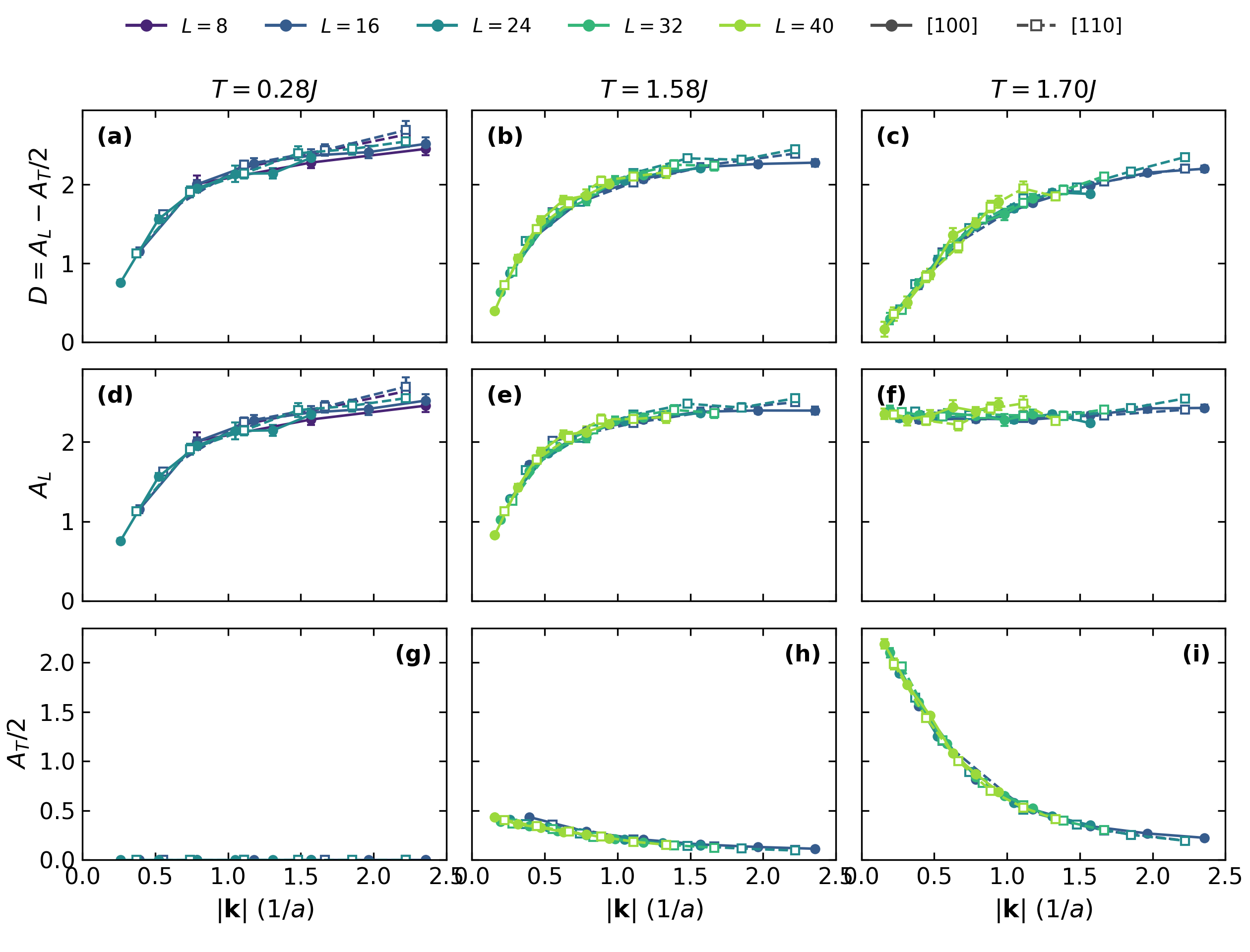}
	\caption{Pinch-point amplitudes near $\mathbf G$ versus $|\mathbf k|$, at $T=0.28J$ (annealing MC, $L=8$--$24$), $1.58J$ and $1.70J$ (PT, $L=16$--$40$), along $[100]$ (filled) and $[110]$ (open); per cell, jackknife errors. (a--c) Contrast $D=A_L-A_T/2$, (d--f) longitudinal amplitude $A_L$, (g--i) transverse amplitude $A_T/2$.}
	\label{fig:sm_pinch}
\end{figure*}

Figure~\ref{fig:sm_Sq_size} shows MC maps of the $(h,k,\frac32)$ plane for $L=8$, $16$, $24$; the $(h,k,0)$ plane, shown in Fig.~\ref{fig:Sq}, contains only the pinch-point (bow-tie) structure, whereas planes with half-integer $l$ show maxima at $\mathbf Q_0+\mathbf G$. Their height is independent of $L$ (log--log slope $\lesssim0.05$, versus $3$ for a Bragg peak) and so is their width (Lorentzian FWHM $\simeq0.13$ in units of $2\pi/a$, a correlation length of about two cells). They rise at $T_c$, where their width drops to its low-temperature value (Fig.~\ref{fig:sm_Sq_T}). Their intensity follows the polarization factor ($0.97$ at $(\frac12,\frac12,\frac32)$ and zero at $(\frac12,\frac12,\frac12)$, where no maximum is seen). Since $m_i=\eta_i\sigma_i$, a uniform $\sigma$ configuration has its weight at $\mathbf Q_0$.
The pinch points themselves are analysed in Fig.~\ref{fig:sm_pinch}. Near $\mathbf G$, with $\mathbf q=\mathbf G+\mathbf k$, the correlation tensor of the arrows splits as $C(\mathbf k)=\frac{A_T}{2}\,\mathbb 1+D\,\hat{\mathbf s}\hat{\mathbf s}^{T}$, with $\hat{\mathbf s}$ the lattice longitudinal direction ($s_\mu=\sin(q_\mu/2)$), $A_L=\hat{\mathbf s}\cdot C\cdot\hat{\mathbf s}$, $A_T=\mathrm{Tr}\,C-A_L$, and $D=A_L-A_T/2$ the amplitude of the angular (bow-tie) part; a singular pinch point requires $D$ to stay finite as $\mathbf k\to0$. At the temperatures studied, the amplitudes collapse onto single curves of $|\mathbf k|$ for all sizes and along $[100]$ and $[110]$, and $D$ decreases towards zero as $|\mathbf k|\to0$: the bow tie has a finite width, i.e., a correlation length of a few lattice spacings. The two mechanisms are distinct. Above $T_c$ the longitudinal amplitude is flat while the transverse one fills in at small $|\mathbf k|$, through the plaquette defects. Below $T_c$ the transverse amplitude is small and vanishes in the ground-state manifold ($T=0.28J$), while the longitudinal one is suppressed at large scales by the star order.
%

\section{Deformation away from the Amp\`ere point}
\label{sm:alpha}

We consider $\mathcal{H}_\alpha=-J\sum_{\rm NN_1}\sigma\sigma'+\alpha J\sum_{\rm NN_2}\sigma\sigma'$, which equals the Amp\`ere Hamiltonian, Eq.~(\ref{eq:H}), up to a constant at $\alpha=1$, and $\mathcal{H}_{\rm Amp\`ere}+(\alpha-1)J\sum_{\rm NN_2}\sigma\sigma'$ otherwise. For $\alpha\neq1$ the extra term lifts the degeneracy of the Amp\`ere manifold. For $\alpha<1$ it favours parallel NN$_2$ spins and selects the uniform-$\sigma$ states: every vertex is a star, with alternating signs on the two sublattices (all-in/all-out order, $f_A=f_B=1$), and there are only two such states. For $\alpha>1$ it favours antiparallel NN$_2$ spins and selects the layered states in which each arrow component depends only on its own coordinate, $m_\mu=m_\mu(r_\mu)$; there are $2^{3L}$ of them, and their star density is close to the random value. Neither phase has an extensive residual entropy: $S_0/N=\ln2/N$ and $\ln2/L^2$, respectively.

\begin{figure*}[t]
	\centering
	\includegraphics[width=0.9\textwidth]{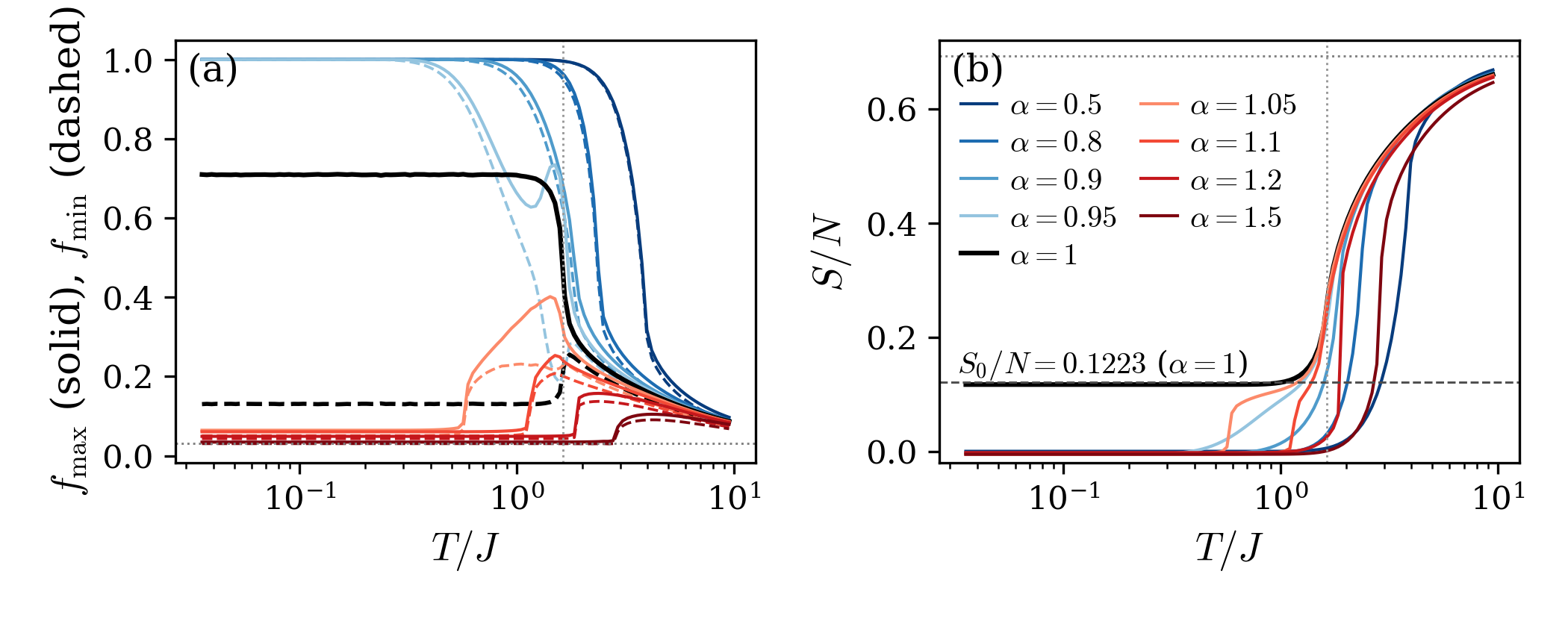}
	\caption{Deformation away from the Amp\`ere point, $H_\alpha=-J\sum_{\rm NN_1}\sigma\sigma'+\alpha J\sum_{\rm NN_2}\sigma\sigma'$ (annealing MC, $L=12$, 20 copies). (a) Star fractions on the sublattice with more ($f_{\max}$, solid) and fewer ($f_{\min}$, dashed) stars; dotted: $1/32$. (b) Entropy per spin from thermodynamic integration of the energy; dashed: $S_0/N=0.1223$ of the Amp\`ere model; dotted: $\ln2$. Vertical dotted lines: $T_c(\alpha=1)$. Only $\alpha=1$ keeps an extensive residual entropy; the slightly lower plateau of $\alpha=1$ ($0.1175$) reflects the coarse temperature grid.}
	\label{fig:sm_alpha}
\end{figure*}

We simulated $\mathcal{H}_\alpha$ with the annealing protocol of Sec.~\ref{sm:methods}, down to $T=0.035J$, for $L=12$ (20 copies per $\alpha$); star moves are no longer free for $\alpha\neq1$ and are accepted with the Metropolis rule. Figure~\ref{fig:sm_alpha}(a) shows the star fractions. For $\alpha<1$ both sublattices fill with stars below a temperature that grows with $1-\alpha$; for $\alpha>1$ the system drops abruptly into the layered states, with few stars on both sublattices. Close to $\alpha=1$ ($\alpha=0.95$, $1.05$) the sublattice star order of the Amp\`ere model appears in an intermediate temperature window, $f_{\max}\neq f_{\min}$, before the low-temperature phase sets in. The entropy [Fig.~\ref{fig:sm_alpha}(b)], obtained by integrating the energy, $S(T)=S(T_0)-\int_T^{T_0}dE/T'$, with $S(T_0)/N=\ln2-(4+2\alpha^2)J^2/(2T_0^2)$ from the exact high-temperature limit of $C_v$, vanishes at low temperature for all $\alpha\neq1$, within the accuracy $\sim0.005$ of the integration. Only $\alpha=1$ retains an extensive residual entropy. 

\section{Correlations of the local pair observable}
\label{sm:pair_correlations}

We define $P(\mathbf R)=\frac{1}{15} \sum_{i<j\in\mathbf R}\sigma_i\sigma_j=\frac{Q_{\mathbf R}^{\,2}-6}{30},$ where the sum runs over the 15 distinct pairs among the six spins incident on vertex $\mathbf R$. Thus $P=1$ for either all-in or all-out stars. We compute the unconnected radial correlation $C_{\mathcal S}(r)=\left\langle\frac{1}{N_{\mathcal S}}\sum_{\mathbf R\in\mathcal S} \frac{1}{N_r}\sum_{\substack{|\boldsymbol{\delta}|=r\\
\mathbf R+\boldsymbol{\delta}\in\mathcal S}} P(\mathbf R)\,P(\mathbf R+\boldsymbol{\delta}) \right\rangle$ Here $N_{\mathcal S}=L^3/2$ is the number of vertices in sublattice $\mathcal S$, and $N_r$ is the number of distinct periodic displacements of minimum-image length $r$ connecting vertices of that sublattice. Angular brackets denote an average over configurations. In each configuration, the sublattice with the larger fraction of $|Q_{\mathbf R}|=6$ vertices is labelled star-rich and the other star-poor, before averaging; ties are treated by averaging both assignments.

\begin{figure*}[t]
\centering
\includegraphics[width=\textwidth]{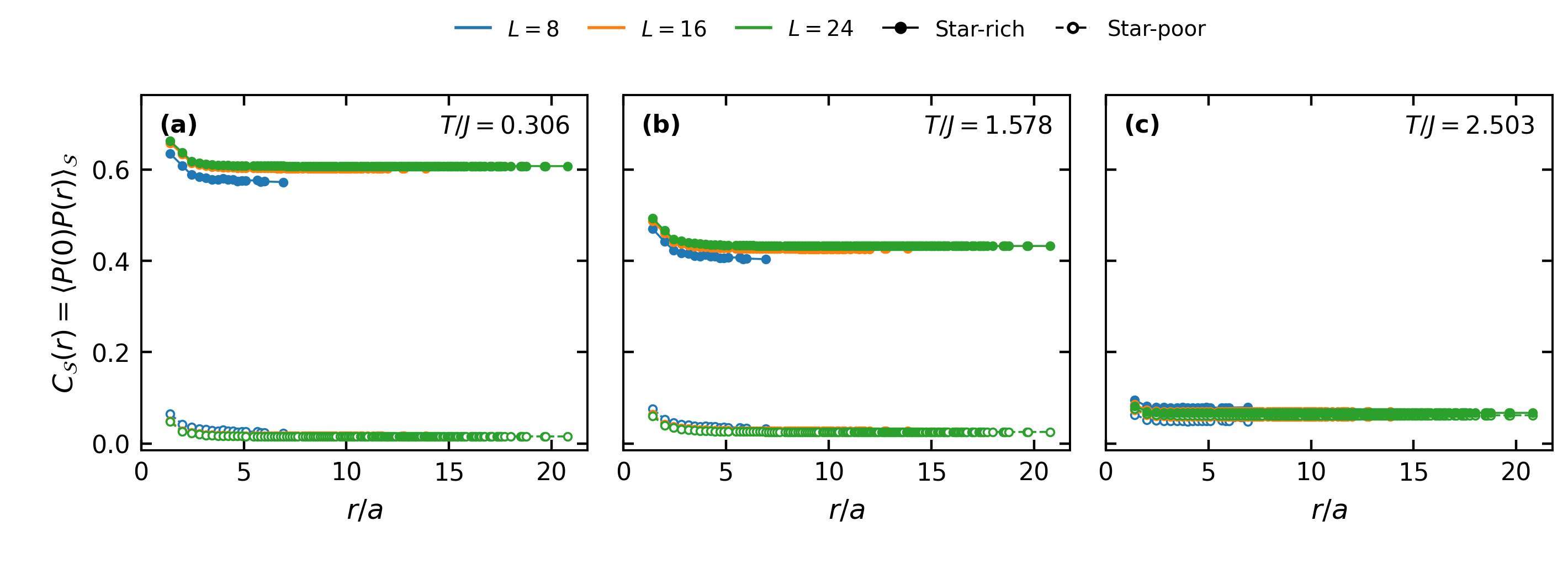}
\caption{Correlations of the normalized local pair observable
$P(r)=(Q_r^2-6)/30$ at (a) $T/J\simeq0.306$, (b) $1.578$, and (c) $2.503$. Blue, orange, and green denote $L=8,16,24$, respectively. Filled (open) circles indicate the star-rich (star-poor) sublattice, identified separately in each configuration; lines guide the eye. Distances are measured in units of the cubic vertex spacing $a$. Error bars are standard errors over independent copies, after averaging stored snapshots within each copy. Below $T_c/J\simeq1.631$, the two sublattices exhibit distinct long-distance plateaus, whereas above $T_c$ their correlations nearly coincide for the larger sizes.}
\label{fig:sm_pair_correlations}
\end{figure*}

Figure~\ref{fig:sm_pair_correlations} shows distinct long-distance plateaus below $T_c$, with close agreement between $L=16$ and $24$, supporting long-range sublattice order in this scalar pair observable. At $T/J\simeq2.503>T_c/J$, the separation decreases with increasing $L$. Since these correlations are unconnected, a nonzero plateau alone does not establish broken sublattice symmetry; the relevant signature is the persistent difference between the two sublattices. Ranking by star fraction
can produce a finite-size difference even in the disordered phase.

\end{document}